\documentclass[11pt,reqno]{amsart}
\usepackage{amsfonts}
\usepackage{amsmath}
\usepackage{amssymb}
\usepackage{color}
\usepackage{graphicx}
\usepackage{xspace}
\usepackage{mathrsfs}
\usepackage{hyperref}
\usepackage[cal=boondoxo]{mathalfa}
\usepackage{forest}
\usepackage{stmaryrd}
\input{xy}
\xyoption{all}

\usepackage{color}
\usepackage{pifont}

\usepackage{pstricks}
\usepackage{pst-grad}
\usepackage{pst-text}
\usepackage{pst-tree}
\usepackage{pst-coil}
\usepackage{pst-node}

\newfont{\indic}{bbmss12}

\def\shuffle{{\scriptscriptstyle \;\sqcup \hspace*{-0.07cm}\sqcup\;}}
\usepackage{xspace}

\def\mapright#1{\smash{\mathop{\lra}\limits^{#1}}}

\def\lra{\longrightarrow}

\def\shuff#1#2{\mathbin{
      \hbox{\vbox{\hbox{\vrule \hskip#2 \vrule height#1 width 0pt}\hrule}\vbox{\hbox{\vrule \hskip#2 \vrule height#1 width 0pt\vrule }\hrule}}}}
\def\shuffl{{\mathchoice{\shuff{7pt}{3.5pt}}{\shuff{6pt}{3pt}}{\shuff{4pt}{2pt}}{\shuff{3pt}{1.5pt}}}}
\def\shuffle{\, \shuffl \,}

\newtheorem{theorem}{Theorem}[section]

\newtheorem{defn}{Definition}

\title[Quantum independences]{Algebraic structures underlying quantum independences : Theory and Applications}

\author[R. Chétrite]{R. Chétrite}
\address{Univ.~C\^ote d'Azur, CNRS,
         		UMR 7351,
         		Parc Valrose,
         		06108 Nice Cedex 02, France.}

\author[F.~Patras]{F.~Patras}
\address{Univ.~C\^ote d'Azur, CNRS,
         		UMR 7351,
         		Parc Valrose,
         		06108 Nice Cedex 02, France.}

\date{\today}

\begin{document}
\maketitle

\section*{Abstract}
The present survey results from the will to reconcile two approaches to quantum probabilities: one rather physical and coming directly from quantum mechanics, the other more algebraic. The second leading idea is to provide a unified picture introducing jointly to several fields of applications, many of which are probably not all familiar (at leat at the same time and in the form we use to present them) to the readers. Lastly, we take the opportunity to present various results obtained recently that use group and bialgebra techniques to handle notions such as cumulants or Wick polynomials in the various noncommutative probability theories. 
\vspace{0.5cm}
\section*{Prelude}
From 2005 to 2008, R.C. did his master and his PhD in Lyon under the supervision of Krzysztof Gawedzki. On the one hand, the subject of master was on the algebraic side of Mathematical Physics: ``Wess-Zumino-Witten-Novikov model, graphs of modular invariants from WZWN with a not simply connected target group''. On the other hand, the subject of  PhD was on the probabilistic side of Mathematical Physics : ``Large deviations and Fluctuation Relations in some models of non-equilibrium systems''. Since his PhD in functional analysis, Krzysztof's scientific path has traversed a large part of the spectrum of mathematical physics.  This survey, at its modest level, tries to adopt this ''metised'' philosophy: juggling between Algebra, Combinatorics, Quantum Physics and Probability. R.C hope that Krzysztof would have enjoyed reading it. I miss him a lot. 
\vspace{0.5cm}

\tableofcontents

\section{Introduction}

This survey aims at putting forward some ideas on the interactions between algebra, algebraic combinatorics and probabilities, insisting on the non-commutative and quantum aspects. Many results are now classical, but their exposition sometimes less so. Some results or constructions are recent, but we do not pretend to originality: if there is any, it will rather reside in the bringing together of various techniques, various points of view whose knowledge is often confined to such and such a field whereas it seems to us to deserve a wider diffusion.

Let's start by noting that the meaning of the expression, ``quantum probabilities", is not entirely fixed in physics and mathematics. Often, it is not simply a question of accounting for the probability calculations of usual quantum mechanics but, more generally, of replacing random variables by observables and the measure by a ``state", that is, formally, a unitary linear form satisfying positivity conditions. The observables themselves live in associative algebras, possibly with involution, or in a $C^\ast$-algebra. When one does not want to insist on the possible applications to quantum physics, one also speaks of noncommutative probabilities. We will use the two terminologies (noncommutative probability, resp. quantum probability) as equivalent in the context of the present article. To take a concrete example, the framework of free probability and its combinatorics is classically relevant to master equations and cumulants for random differential equations, the asymptotic behaviour of random matrices and random matrix processes, or to the study of perturbative expansions the planar sector of quantum chromodynamics (QCD). There is a large literature on the subject. In relation to our forthcoming developments, see e.g. the discussion of master equations and cumulants in \cite{NS} and the bialgebraic approach to planar QFT in \cite{EFP16}. This framework has also recently appeared in relation to  quantum thermalization, respectively
fluctuating quantum many
body systems out-of-equilibrium \cite{Pap22,Hrz22}.

The first step that we put forward next, and which will be one of our guiding threads, is that of a meeting between the combinatorics of words and probability. More concretely, we will develop two applications of this idea, that happen to be closely connected to the Baker-Campbell-Hausdorff problem. We start by explain how classical solutions of the problem, when properly extended to the general framework of preLie algebras, also provide a way to relate various quantum probability theories. 
 The second one aims at promoting evidence for the existence of a universal computation underlying very different phenomena that pertain on one side to group theory, and on the other side to limit theorems in probability. 
We discuss it first in the context of classical probabilities, with the example of card mixings, and then show how it adapts to the more general framework of quantum probabilities. We give in particular a proof of the central limit theorems for coalgebra whose combinatorial content is obtained by mimicking a calculation first done in order to describe power maps in groups  at the level of their algebras of (representative) functions.
In a third step, we explain how quantum probabilities, in their different forms, have recently been the object of new approaches pertaining to algebraic combinatorics, that have given rise to new developments. We discuss in particular the emergence of a ``combinatorics of sentences" naturally extending that of words. To make things concrete, we discuss then these ideas in the context of Wick polynomials. Once again, we put forward (and give evidence for) the idea that there are universal combinatorial and algebraic phenomena underlying these constructions: the construction of Wick polynomials, in classical and quantum probabilities appears for example as the particular case of a very general construction.  We will finally conclude this survey with a complementary step in the direction of quantum probability 
by developping in detail a timely topic --- the 2022 Nobel price in Physics was attributed when we were concluding the redaction of this text --- namely the striking historical experiment stemming from J.S. Bell's ideas, concerning an entangled spin pair, and which leads most physicists to conclude the insufficiency of classical probabilities, or in other words the strict inclusion of classical probabilities in quantum probabilities. 

We underline that Quantum physics is undergoing deep transformations, and new interactions are developing as a high pace: this is well known as far as logic, computer science and computation or crytography are concerned. \textsl{What we hint at here is that transformations and interactions are also developing in another direction, namely group theory, algebraic combinatorics and what is sometimes called the domain of ``higher structures'' that run from structures such as preLie algebras and combinatorial Hopf algebras to category theory or operads} (the latter two topics we don't touch here). We start by introduce now the two most common contexts where the non commutative probabilities theory developed in this survey can be apply :  Quantum mechanics and random matrices. 
\ \ 

\par

\vspace{1cm}

{\bf Notation.} We set $[n]:=\{1,\dots,n\}$. Since we will be using several products, we will denote for example $x^{\times n}$ (resp. $exp^{\times}(x)$) the $n$-th power of $x$ (resp. its exponential) for the product $\times$, etc.
\ \ 

\par
\vspace{1cm}

\subsection{Non commutative probabilities in Quantum Mechanics}

We start with the academic exercise of reformulating classical probabilities in an algebraic form. 
This reformulation will then make \textit{natural} the introduction of quantum probabilities.

\subsubsection*{Algebraic representation of toy model of classical probabilities}

For simplicity, we place ourselves here in the case of a finite state space
 $\Omega=\{ 1,2,...,d \} $ and we note $\varSigma$ the maximal sigma-algebra of the subsets of $\Omega$. Moreover, we consider a probability measure $\mathbb{P}$ on $\Omega$,
which is an application from $\varSigma$ to $\left[0,1\right]$ that satifies
that $\mathbb{P}\left(\Omega\right)=1$, and the other standard asumptions. The triplet $\left(\Omega,\varSigma,\mathbb{P}\right)$
was the essence of the axiomatization of classical probability in
the 30's of the 20th century by Andreï Kolmogorov \cite{kolmogorov}. We will now give from it a unusual
matrix representation.

\begin{itemize}

\item If we usually represent the probability $\mathbb{P}$ by the
row vector $\mathbb{P}\equiv\left(\mathbb{P}\left(1\right),\mathbb{P}(2),...,\mathbb{P}\left(d\right)\right)$,
we will represent it here by the diagonal matrix \textbf{positive
of trace unit} 
\[
\rho\equiv Diag\left(\mathbb{P}(1),\mathbb{P}(2),...,\mathbb{P}(d)\right).
\]
We call ``\textbf{density matrix}'' this matrix representation.


\item Similarly for the indicator function
$1_{A}(i)\equiv\begin{cases}
1\textrm{ if }i\in A\\
0\textrm{ otherwise}
\end{cases}$ 
of an events $A\in\varSigma,$ usually written as the column vector 
$1_{A}\equiv\left(\begin{array}{c}
1_{A}(1)\\
1_{A}\left(2\right)\\
\vdots\\
1_{A}\left(d\right)
\end{array}\right),$
and that we will represent here by the diagonal matrix of the \textbf{ orthogonal projector} \footnote{$\pi_{A}^{2}=\pi_{A}=\pi_{A}^{+}$ where $\pi_{A}^{+}$ is the Hermitian adjoint of $\pi_{A}.$} 
\[
\pi_{A}\equiv Diag\left(1_{A}(1),1_{A}(2),...,1_{A}(d)\right).
\]

In this representation, the probability of event $A$ is obtained as 
\begin{equation}
\mathbb{P}\left(A\right)=\sum_{i\in A}\mathbb{P}(\left\{ i\right\} )=\sum_{i\in\Omega}\mathbb{P}(i)1_{A}(i)=Tr\left(\rho\pi_{A}\right).\label{eq:PEQ}
\end{equation}
 
Similarly, the probability of event $B$ conditional on event
$A$ is obtained as 
\[
\mathbb{P}\left(\left.B\right|A\right)=\frac{\sum_{i\in\Omega}\mathbb{P}(i)1_{A}(i)1_{B}(i)}{\mathbb{P}(A)}=Tr\left(\frac{\pi_{A}\rho\pi_{A}}{Tr\left(\rho\pi_{A}\right)}\pi_{B}\right),
\]
and, what we have obtained here, is the matrix representation of the
conditional probabilities as : 

\begin{equation}
\mathbb{P}\left(\left.\right|A\right)\equiv\frac{\pi_{A}\rho\pi_{A}}{Tr\left(\rho\pi_{A}\right)}.\label{eq:projC}
\end{equation}

\item Finally, a random variable $X:\varOmega\rightarrow\mathbb{R}$,
usually represented by the column vector 
$X\equiv\left(\begin{array}{c}
X(1)\\
X\left(2\right)\\
\vdots\\
X\left(d\right)
\end{array}\right),$ 
is represented here by the \textbf{diagonal Hermitian} matrix
\[
X\equiv Diag\left(X(1),X(2),...,X(d)\right).
\]
This matrix representation of a classical random variable
is then called an \textbf{observable}. The mean of the random variable
(or observable) $X$  under the probability (or density matrix) $\mathbb{P}$
is then given by the relation

\begin{equation}
\mathbb{E}\left(X\right)=\sum_{i\in\Omega}\mathbb{P}(\left\{ i\right\} )X(i)=\sum_{i\in\Omega}\mathbb{P}(i)X(i)=Tr\left(\rho X\right).\label{eq:EC}
\end{equation}

\end{itemize}

In summary, we obtained the ''trace formulas'' 
\[
\begin{cases}
\mathbb{P}\left(A\right)=Tr\left(\rho\pi_{A}\right)\\
\mathbb{P}\left(\left.\right|A\right)\equiv\frac{\pi_{A}\rho\pi_{A}}{Tr\left(\rho\pi_{A}\right)}\\
\mathbb{E}\left(X\right)=Tr\left(\rho X\right)
\end{cases},
\]
which will be the basis of quantum probabilities in the next paragraph.
Let us notice that in the obtained representation, all the
matrices are diagonal: we will say that we have obtained a
commutative algebraic representation. 


\subsubsection*{Quantum mechanics}

The algebraic formalism that we have just presented, derived from the axiomatization of Kolmogorov, is a priori not  sufficient to describe the quantum world. In particular, the reality of objects
and the commutativity hypothesis are put in defect in the quantum world: complex non diagonal matrices are required. This is known since the first
 experiments on interference and polarization measurements. 
The quantum probabilities can be seen as resulting then from the extension of the 
algebraic representation that we have just obtained, but by relaxing these two constraints.

\begin{itemize}

\item More precisely, a quantum system is associated to a
Hilbert space $\mathcal{H}$, which we assume here, for simplicity,
of finite dimension,
 i.e. $\mathcal{H}=\mathbb{C}^{d}.$
 \vspace{0.5cm}
 
\item The state of the system is now described by a \textbf{density matrix} \cite{vN,lan}
 $\rho$ which is a matrix on $\mathcal{H}$, complex, $d-d$, positive,
of unit trace:  i.e. $\rho^{+}=\rho\geq0$ and $Tr\rho=1.$ Here $\rho^{+}$
 denotes the Hermitian adjoint of $\rho.$ This matrix set is
a convex set. Its extremal points (i.e. those which cannot be written
as non-trivial convex combination of density matrices, i.e. states) are the orthogonal projectors of rank $1$, called \textbf{pure states}. These pure states are described by unit vectors
$\psi$ such that

\begin{equation}
\rho=\psi\psi^{+},\label{eq:EP}
\end{equation}
where $\psi$ is called the \textbf{ ket} of the system. 
\vspace{0.5cm}

\item The events associated with quantum systems are then given by
the complex matrices $d-d$ \textbf{of orthogonal projectors} of
$\mathcal{H}$ : 
$\pi^{+}=\pi=\pi^{2}$. 
Their probability is given by the extension
of the formula  (\ref{eq:PEQ}), called Born rules: 

\begin{equation}
\mathbb{P}_{\rho}\left(\pi\right)=Tr\left(\rho\pi\right)=Tr\left(\pi\rho\pi\right).\label{eq:PEQ-1}
\end{equation}

Moreover, the state of the system directly after the event $\pi$
is realized is then given by the extension of the formula (\ref{eq:projC}) : 
\begin{equation}
\left.\rho\right|_{\pi}\equiv\frac{\pi\rho\pi}{Tr\left(\pi\rho\pi\right)}.\label{eq:projC-1}
\end{equation}
 
Let us note that in the case of a pure state (\ref{eq:EP}) $\rho=\psi\psi^{\dagger},$
the probability (\ref{eq:PEQ-1}) takes the form of the \textbf{collapse} of the
wave vector\footnote{Its physical status is still very much debated, in particular because of its instantaneous and non-linear character. Historically, it seems that the first
to have introduced this formula is Werner Heisenberg in 1927 \cite{heisenberg}. }: 
\begin{equation}
\mathbb{P}_{\psi}\left(\pi\right)=\psi^{+}\pi\psi,\label{eq:PEQ-1-2}
\end{equation}
which is quadratic in the wave function $\psi$. This is the basis of
of spectacular interference phenomena \cite{feynman}. 
If in addition the orthogonal projector $\pi$ is of
rank $1,$ i.e. $\pi=\omega\omega^{+}$ with $\omega^{+}\omega=1,$
then we obtain the expression
 
\begin{equation}
\mathbb{P}_{\psi}\left(\pi=\omega\omega^{+}\right)=\left|\psi^{+}\omega\right|^{2},\label{eq:PEQ-1-2-1}
\end{equation}
and the formula (\ref{eq:projC-1}) implies that the pure state is projected
on the pure state $\omega$ after the measurement of $\pi=\omega\omega^{+}$. 

\vspace{0.5cm}

\item Finally, the random variables, now called \textbf{observables},
are given by complex matrices, $d-d$, and \textbf{ Hermitian $X^{+}=X$}. The mean of the \textbf{observable}
$X$ in the state $\rho$ is then given by the extension
of the formula (\ref{eq:projC})  

\begin{equation}
\mathbb{E}_{\rho}\left(X\right)=Tr\left(\rho X\right).\label{eq:EC-1}
\end{equation}

Moreover, the spectral theorem for a Hermitian matrix gives that
$X=\sum_{i}x_{i}\pi_{i}$ where $x_{i}\in\mathbb{R}$ belongs to the spectrum
of $X$ and where the orthogonal projectors verify $\pi_{i}\pi_{j}\sim\delta_{ij}$
and $\sum_{i}\pi_{i}=Id.$ So, if we always suppose true the linearity
of the expectation, the formula (\ref{eq:EC-1}) becomes 
\[
\mathbb{E}_{\rho}\left(X\right)=\sum_{i}x_{i}\mathbb{E}_{\rho}\left(\pi_{i}\right)=\sum_{i}x_{i}Tr\left(\rho\pi_{i}\right)=\sum_{i}x_{i}\mathbb{P}_{\rho}\left(\pi_{i}\right),
\]
where the second equality comes from the relation (\ref{eq:EC-1}) and the
third from the relation (\ref{eq:PEQ-1}). We see that the outcomes
of the random variable $X$ are its eigenvalues $x_{i}$. Then, the \textbf{Von Neumann projective measurement}
of the observable $X$ is the obtaining of the values $x_{i}\in Spectrum\left(X\right)$
with the probabilities (\ref{eq:PEQ-1}) 

\begin{equation}
\mathbb{P}_{\rho}\left(X=x_{i}\right)=Tr\left(\pi_{i}\rho\right),\label{eq:PEQ-1-1}
\end{equation}
and, at the same time, the projection postulate of the density matrix given
by the formula (\ref{eq:projC-1}) 
\begin{equation}
\left.\rho\right|_{\pi_{i}}\equiv\frac{\pi_{i}\rho\pi_{i}}{Tr\left(\pi_{i}\rho\pi_{i}\right)}.\label{eq:projC-1-1}
\end{equation}
\end{itemize}
\vspace{0.5cm}

There are (at least) four physically important points to be made:

\begin{enumerate}
\item The projection postulate (\ref{eq:projC-1},\ref{eq:projC-1-1})  gives an evolution
of the density matrix which is discontinuous in time, random, instantaneous
in time...: several properties which seem strongly unphysical,
and which make it a rule with a strongly debated status. Associated with this,
the physical status of pure states oscillates between a philosophy where
it contains the reality of the world and a philosophy where it is just
the reflection of our knowledge of the world. See the book \cite{laloe} for
much more on this subject.

\item The formulas (\ref{eq:projC-1-1}) required that the results
$x_{i}$ of the measurement be read by the observer.  On the other hand, in
the case where we measure the observable $X$ but we do not read the result 
of the measurement (or we do not know it, because the measurement is made by another
 person), then the density matrix becomes 

\begin{equation}
\sum_{i}Tr\left(\pi_{i}\rho\pi_{i}\right)\frac{\pi_{i}\rho\pi_{i}}{Tr\left(\pi_{i}\rho\pi_{i}\right)}=\sum_{i}\pi_{i}\rho\pi_{i},\label{eq:projC-1-1-1}
\end{equation}
which is different from $\rho,$ except if $\rho$ commutes with all
the $\pi_{i}$. The relaxation of the commutativity constraint
has thus led to the spectacular conclusion that \textbf{an
unread measure has a non-trivial action on the state of the system}.
This action plays an important role for a quantum system in interaction
with its environment, because it is at the basis of the  \textbf{decoherence}\cite{joos} : i.e. the disappearance of the undiagonal coefficients of $\rho$ in a particular basis related to the interaction.


\item 

Matrix non commutativity translates directly into the non
commutativity of the quantum measurement act. 
For example, the probability, in the $\rho$ state, to first measure the observable
$X$ given by the spectral decomposition $X=\sum_{i}x_{i}\pi_{i}$
and to find the value $x_{i}\in Spectrum\left(X\right)$, and then instantly
after measure the observable $Y$ given by the spectral decomposition
$Y=\sum_{j}y_{j}\nu_{j}$ and find the value $y_{j}\in Spectrum\left(Y\right)$,
is obtained by iterating the formulas (\ref{eq:PEQ-1-1},\ref{eq:projC-1-1}):
\begin{eqnarray}
\mathbb{P}_{\rho}\left(X=x_{i};Y=y_{j}\right) & = & Tr\left(\rho\pi_{i}\right)Tr\left(\frac{\pi_{i}\rho\pi_{i}}{Tr\left(\rho\pi_{i}\right)}\nu_{j}\right)=Tr\left(\nu_{j}\pi_{i}\rho\pi_{i}\nu_{j}\right).\label{eq:M1}
\end{eqnarray}

While the probability, in the state $\rho$, to first measure the observable
$Y=\sum_{j}y_{j}\nu_{j}$ and to find the value $y_{j}\in Spectrum\left(Y\right)$,
and then instantaneously after to measure the observable $X=\sum_{i}x_{i}\pi_{i}$ and to find the value $x_{i}\in Spectrum\left(X\right)$, is 

\begin{eqnarray}
\mathbb{P}_{\rho}\left(Y=y_{j};X=x_{i}\right) & = & Tr\left(\rho\nu_{j}\right)Tr\left(\frac{\nu_{j}\rho\nu_{j}}{Tr\left(\rho\nu_{j}\right)}\pi_{i}\right)=Tr\left(\pi_{i}\nu_{j}\rho\nu_{j}\pi_{i}\right).\label{eq:M2}
\end{eqnarray}

In general, these two probabilities are different, except for example if the $\pi_{i}$ and the $\nu_{j}$ are commuting, as we will explicitly illustrate on an example in the next paragraph.

\item  The modern quantum measurement theory generalizes the previous theory
to more general measurements (not necessarily projective) which are necessary to the description of nature, especially for systems observed continuously in time \cite{holevo,wiseman}. 
\end{enumerate}

\vspace{1cm}

\subsection{Noncommutative probabilities in Random Matrix Theory}
\label{RM}

A random matrix is a matrix-valued random variable, that is, a matrix in which entries are random variables. 
Historically, it was first introduced in physics  by Eugene Wigner to model the nuclei of heavy atoms  \cite{Wigner2}. Afterwards, it percolated in most fields of physics and mathematics \cite{Mehta, Akemann, Anderson}.  One of first the goal of random matrix theory was to describe
the distribution of eigenvalues of large random matrices. They give rise
to universal laws quite different from those known for independent random
variables (like the Gaussian law). 
In the eighty, the notion of  freeness introduced by Voiculescu \cite{voiculescu_92} permitted to revisit this asymptotic limit of random
matrices with an algebraic eye, a path which was continued by many people, including Speicher et al \cite{Mingo}.

\newpage

\section{(Quantum) Probability and word combinatorics}
\vspace{1cm}

We are interested in this chapter in the algebraic combinatorics of words and its relations to probability. The topic is classically related to the theory of groups, of their representations, and to the theory of Lie algebras. Classical techniques and results have then been extended to preLie algebras, a process that is still ongoing. Applications include for exemple relations between cumulants in quantum probability --- this is one of the topics we will develop in this section, together with the relevance of certain coalgebraic and bialgebraic constructions.
We should however mention that the same techniques have applications in other fields of physics. For example, similar ideas apply to
 revisiting and extending the scope of Zimmermann's forest formula for the calculation of counterterms in perturbative quantum field theory \cite{Men}.

\subsection{Revisiting Baker-Campbell-Hausdorff}

One of the most classical and well known problem in the  field of the algebraic combinatorics of words, is to express the logarithm of a product of exponentials of matrices. In group theory, this would correspond to the problem of transporting the group law of a Lie group to the Lie algebra of tangent vectors at the unit element of the group. In analysis, this corresponds to the problem, which is roughly equivalent in technical terms, of computing the logarithm of the solution of a linear differential equation not homogeneous in time (i.e. non autonomous) with matrix  or operator coefficients, that is the logarithm of a time-ordered exponential (the so-called Picard or Dyson solution to the differential equation as an infinite sum of iterated integrals). These are respectively the discrete and continuous Baker-Campbell-Hausdorff (BCH) problems. An old but fundational and still relevant reference on the subject is \cite{MPl}. 

The idea is simple: the problem can be formulated in the tensor algebra --- the free associative algebra --- $T(X)$ on an alphabet $X={x_1,\dots,x_n,\dots}$, i.e. in the vector space freely generated by $X^*$, the set of (possibly empty) words $x_{i_1}\dots x_{i_k}$ on $X$. The product is the {\it concatenation} of words:
$$x_{i_1}\dots x_{i_k}\cdot x_{j_1}\dots x_{j_l}:=x_{i_1}\dots x_{i_k} x_{j_1}\dots x_{j_l}.$$
As the exponential $\exp^{\cdot}(x)$ of a letter (or a word $\omega$) is a linear combination of the powers $x^{\cdot n}=x\dots x$ (resp $\omega^{\cdot n}$), the product of two exponentials $\exp^{\cdot}(x)\exp{\cdot}(y)$ is a linear combination of words $x\dots xy\dots y$. The logarithm of this product is a linear combination of all the words on the alphabet $\{x,y\}$. One has therefore to find the coefficient of each word in the development of the logarithm. 

In a more group-theoretical approach (and a better one in view of computations), one looks for the expression of the solution in the free Lie algebra over $X$, that is as a linear combination of iterated Lie brackets of letters, where the Lie bracket in $T(X)$ is defined on words by $[w,w']:=w\cdot w'-w'\cdot w$. The tensor algebra is the enveloping algebra of the free Lie algebra $Lie(X)$ over the alphabet $X$, and the solution can be obtained by considering the action on a product of exponentials of the canonical projection from $T(X)$ to $Lie(X)$. This method is very closely related to the account we will give of convergence to equilibrium phenomena for card shufflings that we will give later on. An account of the classical results on the combinatorics of the BCH problem can be found in Ch. Reutenauer's book, \cite[Chap. 3]{Reutenauer}. A relatively recent contribution on the Lie theoretic formulation of the problem is \cite{Murua}.

There is however, another standard approach to compute the logarithm of the solution of a linear differential equation not homogeneous in time: the Magnus solution to an arbitrary matrix differential equation 
$$X'(t)=A(t)X(t),\ X(0)=1,\ X(t)=\exp(\omega(t))$$
where
$$\omega(t)=\int\limits_0^t\frac{ad_{\omega(u)}}{\exp(ad_{\omega(u)})-1}A(u)du,$$
where the adjoint action is defined by $ad_{M}N:= [M,N]$. 
It happens that the bilinear operation on time-dependent matrices (normalized so that $M(0)=N(0)=0$),
$$M(t)\{N(t)\}:=\int\limits_0^t[N(u),M'(u)]du,$$
is a preLie product, that is it satisfies the identity:

$(M(t)\{N(t)\})\{P(t)\}-M(t)\{N(t)\{P(t)\}\}$
$$=(M(t)\{P(t)\})\{N(t)\}-M(t)\{N(t)\{P(t)\}\}.$$
This identity defines more generally the notion of (right) preLie algebra (a preLie algebra is a vector space $L$ equipped with a bilinear map $x\{y\}$ that satisfies this identity). The notion was introduced and rediscovered independently several times by Lazard, Gerstenhaber (who proposed the name preLie), Vinberg and Agrachev-Gamkrelidze \cite{Laz,Ger,Vin,AG}. A survey of the theory with details, proofs, together with bibliographical and historical indications can be found in \cite[Chap. 6]{CP21}. 

For various reasons \cite{LM,GO}, it is then convenient to introduce recursively the family indexed by positive integers $n$ of the so-called symmetric brace operations. They are linear in the left argument (denoted $v$ below) and symmetric multilinear operations on the right ones (denoted $w_1,\dots,w_n$ below) and are defined implicitely by
\begin{equation}\label{productbracefla}
v\{w_1,\dots,w_n\}:=(v\{w_1,\dots,w_{n-1}\})\{w_n\}-\sum\limits_{i=1}^{n-1}v\{w_1,\dots,w_i\{w_n\},\dots,w_{n-1}\}.
\end{equation}
These operations allow among others to define an associative product $\ast$ on the polynomial algebra over a preLie algebra $L$: for $a_1,\dots,a_l$ and $b_1,\dots,b_m$ in $L$,
\begin{equation}\label{astproduct}
(a_1...a_l)\ast (b_1...b_m)=\sum\limits_f B_0(a_1\{B_1\})...(a_l\{B_l\}),
\end{equation}
where the sum is over all maps $f$ from $\{1,...,m\}$ to $\{0,...,l\}$ and $B_i:=\prod_{j\in f^{-1}(i)} b_j$. 
For example, in low degrees:
$$
a\ast b=ab+a\{ b\},
$$
$$a_1a_2\ast b=a_1a_2b+(a_1\{b\})a_2+a_1(a_2\{ b\}),$$
$$a\ast b_1b_2=ab_1b_2+b_1(a\{ b_2\})+b_2(a\{ b_1\})+a\{b_1b_2\}$$
$$=ab_1b_2+b_1(a\{b_2\})+b_2(a\{ b_1\})+a\{b_1\ast b_2 -b_1\{ b_2\}\}$$
$$=ab_1b_2+b_1(a\{b_2\})+b_2(a\{ b_1\})+(a\{ b_1\})\{ b_2\} -a\{b_1\{ b_2\}\}.$$

Agrachev and Gamkrelidze \cite{AG}, driven by problems in the theory of dynamical systems and control theory, introduced in this general setting the Magnus operator. This is an automorphism of $L$ (as a set), which satisfies the fixed point equation: $$\forall v\in V, \Omega(v)=v\left\lbrace \frac{\Omega(v)}{\exp^{\ast \Omega(v)}-1} \right\rbrace =v+\sum\limits_{n>0}\frac{B_n}{n!} v\{\Omega^{\ast n}(v)\},$$
where the $B_n$ are the Bernoulli numbers.
The first terms are
$$\Omega(v)=v-\frac{1}{2}v\triangleleft v+\frac{1}{4}v\triangleleft (v\triangleleft v)+\frac{1}{12}(v\triangleleft v)\triangleleft v+\dots .$$
For such an algebra, the bracket operation $$[x,y]:=x\{y\}-y\{x\}$$
can be shown to be a standard Lie bracket (it is antisymmetrical and satisfies the Jacobi identity), 
and the Magnus solution can be understood (abstractly) as lifting the BCH problem from Lie algebras to preLie algebras. This idea was developed systematically in \cite{CP13}. Namely, it was shown in that article that the Magnus operator identifies with the action on a suitable exponential of the canonical projection of the enveloping algebra of a preLie algebra $L$ to $L$. This ``suitable" exponential is a generalization of the time-ordered exponential and the action of the canonical projection is a generalization of the logarithm, see also \cite{bauer}, \cite[Chap. 6]{CP21} and \cite{efp22}.

\subsection{An application in quantum probability}

Our interest will focus in the present section on the application of these ideas to quantum probability. Let us be clear and make a general statement: we do not claim here that a notion such as free independence can allow to solve the aporias of quantum mechanics that will be analysed later in this survey. What we want to show different: namely that by taking seriously the question of generalizing classical probabilities into ``noncommutative" probabilities, a whole set of possibilities opens up, which has led to the development of a field where discrete mathematics plays an essential role. And these new ideas {\it could} bring new tools to deepen {\it also} our understanding of quantum mechanics.

Let us recall first the definition of classical and tensor cumulants, that will be useful also later on in this article. Let us fix $A$ a (commutative) algebra of complex-valued random variables with expectation map $\mathbb E$ and moments at all orders. By independence we will mean polynomial independence: $X$ and $Y$ are called independent if and only if $\mathbb E(X^nY^m)=\mathbb E(X^n)\mathbb E(Y^m)$ for all pairs of positive integers $(n,m)$. 

The quantum analog is a noncommutative probability space, defined as a pair $(\mathcal A,\varphi)$ where $\mathcal A$ is an associative unital algebra over $\mathbb C$ and $\varphi$ a unital linear form. One often requires $\mathcal A$ and $\varphi$ to satisfy further conditions (the former can be a $C^\ast$-algebra, the latter satisfy positivity conditions...), but they do not need to be taken into account for our purposes. Elements in $\mathcal A$ are called noncommutative random variables, or simply random variables. Typical examples of such algebras are : 
\begin{itemize}
\item Algebras of random matrices (see pargraphe \ref{RM}) with the state operator $\mathbb{E}\circ Tr$ (``expectation of the trace''--which is normalized by the dimension of the space to satisfy the condition of unitarity) 
 \item Operator algebras of quantum physics, with $\varphi$  given by (\ref{eq:EC}) $\varphi(X)=Tr\left(\rho X\right)$
 or $\varphi(X)= \psi^{+} X \psi$ for a pure state \eqref{eq:EP} $\rho=\psi\psi^{+},$ and  $\varphi(X)=\emptyset^{+} X\emptyset $ in the particular case where the ket $\psi$ is the fundamental state of the system (which is often the vacuum, hence our notation). 
\end{itemize}

{\it Tensor independence} is defined similarly as in the commutative case. It is also the most common notion of independence in quantum probability --- the one at play for example in Bell's experiment that we will analyze later on. Given nonnegative integers $n_1,\dots,n_{2k}$ where $n_2,\dots,n_{2k-1}$ are positive, $X$ and $Y$ in $\mathcal A$ are tensor independent if and only if 
$$\varphi(X^{n_1}Y^{n_2}\dots X^{n_{2k-1}}Y^{n_{2k}})=\varphi(X^{n_1+n_3+\dots+n_{2k-1}})\varphi(Y^{n_2+\dots+n_{2k}}),$$
which happens typically in quantum physics when $\mathcal A=\mathcal A_1\otimes\mathcal A_2$ and $\varphi=\varphi_1\otimes\varphi_2$, that is, in practice, when considering observables associated to the tensor product of two Hilbert spaces.

Let now $X$ be a real random variable with moments at all orders, $m_n := {\mathbb E}(X^n)$. It is associated the exponential generating series of momenta
\begin{equation}
  M(t):={\mathbb E}(\exp({tX})) = 1+ \sum_{n > 0}m_n\frac{t^n}{n!}.
\end{equation}
In the multivariate case (that is for a family $(X_1,\dotsc,X_p)$ of random variables) one gets similarly the multivariate generating series
$$
    M(t_1,\dotsc,t_p) := \mathbb E(\exp({t_1X_1+\dotsb+t_pX_p}) ).
$$

The exponential generating series $K(t)$ of cumulants $c_n$ (written respectively $K^X(t)$ and $c_n^X$ when the dependency on $X$ has to be taken into account explicitely)
\begin{equation*}
  K(t) :=  \sum_{n>0} c_n \frac{t^n}{n!}
\end{equation*}
is then given by
\begin{equation}
  M(t)=\exp ( K(t) ),
\end{equation}
with, equivalently,
\begin{equation}
  m_n=\sum_{\pi\in P([n])}\prod_{B_i\in\pi}c_{|B_i|}.
\label{eq:cmcrel}
\end{equation}
where $P(E)$ denotes the set of (unordered) partitions $\pi:=\{B_1,\ldots,B_l\}$ of a set $E$.
One gets that if the random variables $X$ and $Y$ are independent, 
$$K^{X+Y}(t)=\log({\mathbb E}(\exp({t(X+Y)}))$$
$$=\log({\mathbb E}(\exp({tX})){\mathbb E}(\exp({tY})))=K^X(t)+K^Y(t).$$

To handle the multivariate case, we fix the following notation: given a family $\{f_n:A^n\to\mathbb{C}\}_{n\geq1}$ of functions and $\pi=\{\pi_1,\dots,\pi_k\}$ a partition of $[n]$, we write
\begin{equation}
    f_\pi(a_1,\ldots,a_n) := \prod_{i\leq k} f_{|\pi_i|}(a_1,\ldots,a_n|\pi_i),
\end{equation}
where for a subset $S$ of $[n]$, $f_{|S|}(a_1,\ldots,a_n|S) := f_{p}(a_{i_1},\ldots,a_{i_p})$ and $|S|=p$ if $S = \{i_1<\cdots< i_p\}$ (notice that with the same convention, $|\pi|=k$). The same notation is used when $A$ is replaced by $\mathcal A$.

The multilinear cumulants maps can then be defined implicitely by
$$\mathbb E(a_1\cdots a_n) = \sum_{\pi\in P([n])} c_{\pi}(a_1,\ldots,a_n).$$ It can be shown easily that they characterize independence: for example, $X$ and $Y$ are independent if and only if all mixed cumulants (that is expressions such as $c_5(X,Y,X,Y,Y)$ with at least one copy of $X$ and of $Y$) vanish.

Recall now that a non-crossing partition of $[n]$ is a partition of $[n]$ such that there are no distinct blocks $\pi_i,\pi_j$ and $a<b<c<d\in [n]$ such that $a,c\in \pi_i$, $b,d\in \pi_j$. The set of non-crossing partitions of $[n]$ is denoted by $NC([n])$. One can define an order on the blocks of a non-crossing partition by $\pi_i\geq\pi_j$ iff $\exists a,b\in \pi_j,c\in \pi_i$ such that $a<c<b$. This poset is a forest: the connected components of its Cayley graph are trees, that is the corresponding subposets have a unique minimal element and are without loops (they are simply connected graphs). The forest associated to $\pi$ is a tree if and only if the partition is irreducible, that is if $1$ and $n$ belong to the same block. The set of irreducible non-crossing partitions of $[n]$ is denoted $NC_{ir}([n])$. An interval partition of $[n]$ (resp., in this survey, of an arbitrary finite subset $S$ of the positive integers) is a partition whose blocks are of the form $\{i,i+1,\ldots,i+j\}$, for some $1\leq i\leq n$ and $0\leq j\leq n-i$. The set of interval partitions of $[n]$ (resp. $S$) is denoted by ${I}([n])$ (resp. $I([S])$). 

Given a tree $t$, the coefficient $t!$ is called the tree factorial. It is recursively defined on trees by $t! = 1$ if $t$ is the single-vertex tree, and if $t$  is obtained by grafting the subtrees $s_1\cdots s_m$ to its root, then $t! = |t|s_1! \cdots s_m!$. For a forest $f$ (a disjoint union) of trees $t_1,\ldots,t_n$, $f!: = t_1!\cdots t_n!$.

In the noncommutative case, \textit{tensor}, \textit{free} \cite{Spe94}, \textit{Boolean} \cite{speicher_97c}, \textit{and monotone} \cite{hasebesaigo_11} \textit{cumulants} are then, respectively, the families of multilinear maps $t_n,r_n,b_n$ and $h_n$ from $\mathcal A^n$ to $\mathbb{C}$ implicitly defined by the equations 
\begin{eqnarray}
\varphi(a_1\cdots a_n) &=& \sum_{\pi\in P([n])} t_{\pi}(a_1,\ldots,a_n),
\\\varphi(a_1\cdots a_n) &=& \sum_{\pi\in NC([n])} r_{\pi}(a_1,\ldots,a_n),
\\\varphi(a_1\cdots a_n) &=& \sum_{\pi\in {I}([n])}  b_{\pi}(a_1,\ldots,a_n),
\\\varphi(a_1\cdots a_n) &=& \sum_{\pi\in NC([n])} \frac{1}{t(\pi)!} h_{\pi}(a_1,\ldots,a_n).
\end{eqnarray}

One can directly relate these families of cumulants to the corresponding notions of independence in quantum probability. For example, two random variables $X,Y$ are freely independent if and only if all their mixed cumulants vanish. Relating monotone cumulants to monotone independence is slightly more involved, this is linked to the fact that monotone independence of $X$ and $Y$ is not a symmetric relation --- this leads to defining antimonotone independence of $X$ and $Y$ as the monotone independence of $Y$ and $X$.  An important result for our purposes is that these notions of independence are the only ones satisfying natural universal properties in quantum probability, as shown by Ben Ghorbal, Muraki, Sch\"urmann and Speicher \cite{Mura03}.

A natural question is then to relate the different notions of independence through their cumulant functions. Formulas were obtained in \cite{Lehner,lehner-etal_15}  using the combinatorial properties of lattices of set partitions. For example,
$$b_n(a_1,\ldots,a_n) = \sum_{\pi\in NC_{{ir}}([n])} r_\pi(a_1,\ldots,a_n).$$
Using a different approach, missing formulas were obtained recently in \cite{cepp} such as
$$h_n(a_1,\ldots,a_n) = \sum_{\pi\in NC_{ir}([n])} (-1)^{|\pi|-1}\omega(t(\pi)) r_\pi(a_1,\ldots,a_n),$$
where $|\pi|$ is the number of blocks of the partition and where, given a tree $t$ with $p$ vertices viewed as an ordered set, $\omega(t)$ is defined as 
\begin{equation}
    \omega(t) := \sum_{k=1}^{p} \frac{(-1)^{k-1}}{k}\omega_k(t).
\end{equation}
where, for any integer $0 < k \leq p$, we denote by $\omega_k(t)$ the number of surjective, strictly order preserving maps $f : t \to [k]$. 

This other approach is based on arguments based on shuffle products of sentences (that will be explained in the last section of this survey) and on preLie algebra formulas. 
Consider the non-unital tensor algebra $T_+(\mathcal A) := \bigoplus_{n>0} \mathcal A^{\otimes n}$ and denote the graded dual by $ \mathfrak{g}:= \bigoplus_{n>0} (\mathcal A^{\otimes n})^\ast.$ Elements $a_1\otimes\dots\otimes a_n$ in $\mathcal A^{\otimes n}$ are written using the word notation $a_1\dots a_n$
Tensor, free, Boolean and monotone cumulants are families of multilinear forms on $\mathcal A$ and can therefore be viewed as elements $t,r,b,h$ of $ \mathfrak{g} $.
The following property is central to the algebraic approach to quantum probability: 
Let $\alpha,\beta \in \mathfrak{g}$ and $w \in T_+(A)$, then the identity
\begin{equation}
	\beta\{\alpha\} (w)
	:= - \sum_{w_1\cdot w_2\cdot w_3 = w}
	 \beta(w_1\cdot w_3)\alpha(w_2),
\end{equation}	 where the sum is over rewritings of $w$ as a concatenation product of nonempty subwords,
defines a preLie algebra structure on $\mathfrak{g}$.

One can then show that monotone, free and Boolean cumulants, $\rho, \kappa, \beta \in \mathfrak{g}$ are related in terms of the Magnus operator
\begin{equation}
	h 	= \Omega(r) 
		= -\Omega(-b).
\end{equation}

The proof of these identities requires shuffle algebra arguments as they will be hinted at in the last section of this survey. For details we refer the reader to \cite{ebrahimipatras_15,EP16,EFP18,EFP19}. However, from the knowledge of these identities many calculations can be performed that depend only on properties of preLie algebras and their enveloping algebras --- as illustrated by the various explicit combinatorial identities connecting free, Boolean and monotone theories, that can be deduced almost automatically from the analysis of the Magnus operator and its inverse.

\subsection{Shuffles and Markov chains}

A key combinatorial notion for the problems we address in this survey is that of shuffle and the dual notion of descent. We will approach it through a problem of statistics where one can find, in a rather surprising way, a whole set of key ideas of the algebraic combinatorics of words but also arguments that we will find later again appearing in the analysis of central limit theorems.

The example goes back to one of the founding texts of probability, Poincaré's {\it Calcul des probabilit{\'e}s}, from 1912.
We find there what is probably the first idea of Markov chains and their convergence to equilibrium. Let us quote Poincaré:
\begin{quote}And
to take a cruder example [of convergence to a random distribution], this is also what happens
when we shuffle cards. At each move, the cards
undergo a permutation (analogous to those studied
in the theory of substitutions). Which one
will happen? The probability that it will be such permutation
[for example, the one that brings to rank $n$ the card that occupied
the rank $\phi(n)$ before the permutation], this probability,
I say, depends on the habits of the player. But if this player
shuffles the cards long enough, there will be a large number of
successive permutations and the final order which will result from it
will be governed only by chance. I mean that all possible orders
will be equally probable. It is to the great number of successive permutations, that is to say
to the complexity of the phenomenon, that this result is due. \cite[p. 9]{Poincare}
\end{quote}

Poincaré's idea reads as follows \cite[p. 13]{Poincare}: a player has habits and, each time he shuffles the cards, he will reorder them according to a certain law of probability. If the cards are numbered in the order in which they appear in the initial deck, say a three-cards deck of aces of clubs, hearts and spades, then each possible configuration, 123, 132, 213, 231, 312, 321, will come out with a certain probability, which depends on the player (but not on which card is in position 1, 2 and 3 in the initial pack). By iterating the process a sufficient number of times, one will approach the uniform probability (1/6 for each configuration), except if the operation is not sufficiently mixing, which is the case if the player is satisfied for example by exchanging only the first and the second card with a certain probability, or otherwise to leave the deck unchanged. Poincar\'e develops the calculation at the end of the book (chapter XVI), with the tools of linear algebra. The spectral analysis of the transition matrix between states shows that, apart from degenerate situations, 1 is the highest modulus eigenvalue and is isolated. The associated eigenvector gives the limit state to which the process converges (the uniform law on all possible orders for a set of $n$ distinct cards). The minimal difference in modulus with the other eigenvalues determines the (exponential) speed of convergence towards the equilibrium. The text is brief, luminous, brilliant: a  masterpiece.

With the formalism we have sketched, fixing an operation of mixing $n$ cards is equivalent to giving a probability law on the group of permutations of $n$ objects, $S_n$, i.e. a family of weights $(p_\sigma)_{S_n}$, positive or null and of sum 1. The weight $p_\sigma$ is the probability that the permutation $\sigma\in S_n$ is applied to the deck during the shuffle, so the transition probability of going from the $x$ configuration to the $\sigma x$ configuration is $p_\sigma$.
Poincar\'e's analysis works regardless of the method of shuffling the cards, i.e. of the collection of  weights $p_\sigma$. The degenerate cases correspond to the situation where the permutations of nonzero weight generate a strict subgroup of $S_n$. 

After Poincaré, probabilists, first and foremost P. Diaconis and his collaborators, were mainly interested in describing precisely the asymptotics of mixings under specific (and realistic) hypotheses on the $p_\sigma$ \cite{aldous,persi}. 
Let us give as an example two possible mixing strategies --- there are obviously many others, we will see later how to code them algebraically. The first one consists in taking the first card of the deck and inserting it equiprobably in all possible positions in the remaining deck. It is clear that this method is non-degenerate, but that it will converge very slowly to the equilibrium (the uniform law). The second method, (which we will call here ``standard mixing") consists in cutting in all possible ways, and then shuffling the two resulting decks. Shuffling means here that we mix the two decks in such a way as to obtain all the total orders on the cards compatible with the total orders on the two decks obtained after cutting (we will give a formal definition later).
This is roughly the method that amateur players spontaneously use. Here again, the method is clearly non-degenerate, and one guesses that it will converge much more quickly to the equilibrium.

Let us take the example of a deck of two cards $A,B$ (we follow hereafter the presentation of \cite{pp}). The word $AB$ denotes the deck where $A$ is above $B$. Starting from $AB$, the first method gives two possible operations, which are equiprobable
$$AB\longmapsto (A,B)\longmapsto AB,BA,$$
where we note $(A,B)$ the fact of having, after having taken the first card, a deck formed by the card $A$ and another one by the card $B$. In this (trivial) case, we obviously converge to the equilibrium already after the first shuffling operation.

The second method gives three cuts (represented again by pairs whose elements correspond to the two obtained decks) and in all four possible operations, three of which preserve the initial configuration
$$AB\longmapsto (AB,\emptyset)\longmapsto AB$$
$$AB\longmapsto (A,B)\longmapsto AB,BA$$
$$AB\longmapsto (\emptyset,AB)\longmapsto AB,$$
If we suppose that all the paths are equiprobable, we obtain $AB$ with probability $3/4$ and $BA$ with probability $1/4$.

The mixing operation is thus represented by a Markovian transition matrix:
\begin{equation*}
M = 
\begin{pmatrix}
3/4 & 1/4 \\
1/4 & 3/4
\end{pmatrix}
\end{equation*}
such that, if we start from a distribution $P$ (a probability law on the possible orders): $P(AB)=p, P(BA)=1-p$, we obtain after a standard mixture the distribution $Q$, $Q(AB)=1/2\ p+1/4,Q(BA)=
3/4-1/2\ p$ since, by agreeing to represent the distributions by line vectors and to make $M$ operate on the right :
\begin{equation*}
\begin{pmatrix}
p, &
1-p
\end{pmatrix}M=
\begin{pmatrix}
p, &
1-p
\end{pmatrix}\begin{pmatrix}
3/4 & 1/4 \\
1/4 & 3/4
\end{pmatrix}
=
\begin{pmatrix}
1/2\ p+1/4,&
3/4-1/2\ p . 
\end{pmatrix}
\end{equation*}
The key idea of Poincar\'e is that the time evolution of the probability law on the configurations is governed by the spectral analysis of the matrix $M$. Its dominant eigenvalue is $1$, associated with the uniform law $U(AB)=1/2,\ U(BA)=1/2$. The second eigenvalue is $1/2$ with associated eigenvector \begin{equation*}
\begin{pmatrix}
1/2,&
-1/2
\end{pmatrix}.
\end{equation*}
In general, for a pack of $N$ cards, the spectrum of the Markovian matrix describing standard mixings is $1,1/2,\dots ,1/2^{N-1}$, the uniform law corresponding to the eigenvalue $1$. We will see later how to obtain these results.

Returning to the situation where we start from the $AB$ deck to which we apply $k$ standard mixings, we obtain
\begin{equation*}
\begin{pmatrix}
1, &
0
\end{pmatrix} M^k = 
\begin{pmatrix}
1, &
0
\end{pmatrix}\begin{pmatrix}
3/4 & 1/4 \\
1/4 & 3/4
\end{pmatrix}^k=\begin{pmatrix}
1/2,&
1/2
\end{pmatrix}+1/2^k
\begin{pmatrix}
1/2,&
-1/2
\end{pmatrix}.
\end{equation*}
The obtained distribution is $Q_k(AB) = 1/ 2 +1/{2^{k+1}}$ and $Q_k(BA) = 1/ 2 -{1}/{2^{k+1}}$,
with exponential convergence to the uniform distribution. 
This phenomenon is the convergence to equilibrium  of Markov chains, it applies in a wide variety of situations. The speed of convergence is measured by the ``spectral hole'': the minimal distance between $1$ and the other eigenvalues, here $\frac{1}{2}$.

\vspace{1cm}
\subsection{Coalgebras and Hopf algebras}

We will briefly explain now how to encode these phenomena with group-theoretic methods, very close to those used in algebraic combinatorics of words. The key notion for everything that follows in the chapter is that of coalgebra, dual to that of algebra, and that of bigebra, which mixes algebra and coalgebra structures.
A coproduct on a vector space $C$ is a linear application 
$$\Delta : C\to C\otimes C=C^{\otimes 2}.$$
If $C$ has a basis $(b_i)_{i}$, the tensor product $C\otimes C$ has as a basis $(b_i\otimes b_j)_{i,j}$, and we cannot in general write elements such as $b_1\otimes b_2+b_3\otimes b_4$, in the form $v\otimes v'$ for $v,v'\in C$. This problem is at the heart of quantum entanglement, which will be discussed again later in connection with Bell's inequalities. For example: the pure state of a pair of Bell spins gives an example of an entangled state (\ref{etatpur}). However, to describe the action of the coproduct, we use the Sweedler notation $$\Delta(c)=:c^{(1)}\otimes c^{(2)}.$$
The notation is abusive (because of the phenomenon of entanglement it is impossible in general to write $\Delta(c)$ as a genuine tensor product $v\otimes v'$, see \ref{bellpair}), but its use does not pose any problems in practice\label{sweedler}.

The coproduct is coassociative if we have the equality between applications from $C$ to $C^{\otimes 3}$
\begin{equation}{\label{coass}}
 (\Delta\otimes Id)\circ \Delta = (Id\otimes\Delta )\circ \Delta.
\end{equation}
When the coproduct is coassociative, there is a unique application from $C$ to $C^{\otimes n}$ obtained by iterated composition of $n-1$ coproducts. We write it $\Delta_n$, it is defined by $\Delta_2:=\Delta$, $$\Delta_n:=(\Delta\otimes Id^{\otimes n-2})\circ \Delta_{n-1}.$$
The coproduct is cocommutative if $$T\circ\Delta=\Delta ,$$
where $T(a\otimes b):=b\otimes a$. In Sweedler notation, the coproduct is cocommutative if
$$c^{(1)}\otimes c^{(2)}=c^{(2)}\otimes c^{(1)}.$$
The dual notion of unit for an algebra is that of counit: a linear application $\varepsilon$ from $C$ to the ground field, taken most often to be $\mathbb C$ in this article, such that
\begin{equation}\label{counit}
(Id\otimes\varepsilon)\circ \Delta=Id=(\varepsilon\otimes Id)\circ\Delta ,
\end{equation}
where ${\mathbb C}\otimes C\cong C\otimes {\mathbb C}\cong C$.

\begin{defn}
A coalgebra is a vector space $C$ equipped with a coassociative coproduct $\Delta$ and a counit $\varepsilon$. It is cocommutative if $\Delta$ is cocommutative.
\end{defn}

A classical example of a coalgebra is the coalgebra of a finite partially ordered set $P$. We take for $C$ the vector space generated by the pairs $(x,y)$ with $x\leq y$. The coproduct
$$\Delta(x,y):=\sum\limits_{x,z}(x,z)\otimes (z,y);\varepsilon(x,y)=\delta_{x=y}$$
then defines a coalgebra structure on $C$. This coalgebra structure is useful to encode M\"obius inversion, a useful tool to handle computations in quantum probability due to the fact that cumulants in the various theories are defined as sums indexed by various families of partitions that have a poset (and even a lattice) structure. See for example the seminal book by Nica and Speicher for insights on the use of lattice techniques in free probability \cite{nicaspeicher_06}.

The relevant coalgebra to study mixings of cards is the one describing cuts: it is the dual application to the concatenation product of words. Formally, 
$$\Delta_{c}:T(X)\to T(X)\otimes T(X),$$
$$\Delta_{c}(y_1\dots y_n)=\sum\limits_{i=0}^n y_1\dots y_i\otimes y_{i+1}\dots y_n,$$
where the $y_i$ are letters of $X$.

\begin{defn}
A bialgebra is a 5-tuple $(B,m,\eta,\Delta,\varepsilon)$ where :
\begin{itemize}
\item $(B,m,\eta)$ is an algebra (associative, with unit application $\eta : {\mathbb C}\to B$)
\item $(B,\Delta,\varepsilon)$ is a coalgebra (with counit $\varepsilon: B\to {\mathbb C}$)
\item The coproduct $\Delta$ and the counit $\varepsilon$ are morphisms of algebras. Equivalently, the product $m$ and the unit $\eta$ are morphisms of coalgebras.
\end{itemize}
\end{defn}
A key operation in bialgebra theory is the convolution product: if $f$ and $g$ are two linear endomorphisms of $B$, we define their convolution product by 
$$f\ast g:B\mapright{\Delta}B\otimes B\mapright{f\otimes g}B\otimes B\mapright{m} B.$$
One easily verifies that $\ast$ provides $End(B)$ with an associative algebra structure with unit $\nu:=\eta\circ\varepsilon$.
Note for later use that the definition can be adapted to define the convolution product of two linear forms $\alpha, \beta$ on a coalgebra $C$ by
$$\alpha\ast \beta:C\mapright{\Delta}C\otimes C\mapright{\alpha\otimes \beta}{\mathbb C}\otimes {\mathbb C}=\mathbb C.$$
We skip details. The reader can find in \cite{CP21} a detailed exposition of the theory of coalgebras, bialgebras, and their applications.

The relevant bialgebra for studying mixings of cards is obtained by adding to the coproduct the shuffle product, defined inductively by using its decomposition in two ``half-shuffles"
$$y_1\dots y_n\shuffle z_1\dots z_m:=y_1\cdot(y_2\dots y_n\shuffle z_1\dots z_m)+ z_1\cdot (y_1\dots y_n\shuffle z_2\dots z_m).$$
One can easily convince oneself that this product models the standard shuffle of two decks of cards as described earlier.

Let us now explain how to encode the different possible shuffles with the bigebra structure of $T(X_n)$, where $X_n=\{x_1,\dots,x_n\}$ now denotes a set of $n$ cards, assumed to be distinct for convenience. Let $X=\bigcup\limits_{n}X_n$ and embed $T(X_n)$ into $T(X)$.
A probability law on the possible orders of the cards in $X_n$ is coded by a linear combination with positive coefficients of sum 1 of the words without repetitions on the letters of $X_n$ and of length $n$. We take till the end of this section the rationals $\mathbb Q$ as the ground field.
Let $p_k$ be the identity application on the words of length $k$ of $T(X)$ and the null application on the other words. One can show (see \cite[Chap. 5]{CP21}) that 
\begin{enumerate}
\item the convolution algebra $\mathcal D$ generated by the graded projections $p_k$ (a subalgebra of the convolution algebra of endomorphisms of $T(X)$) is a free associative algebra,
\item that it is stable by composition (of endomorphisms of $T(X)$),
\item that it has itself a bialgebra structure, obtained by setting
$$\Delta(p_n)=\sum\limits_{i=0}^np_i\otimes p_{n-i},$$
\item that its action on words is described by linear combinations of permutations (a convolution product of $p_i$ acts on words of length $k$ as a linear combination of elements of the $k$-th symmetric group $S_k$).
\end{enumerate}
The opposite algebra $\mathcal D^{op}$ is called the descent algebra because the description of the elements of $\mathcal D$ and $\mathcal D^{op}$ seen as linear combinations of permutations relies on the statistics of the descents of permutations: for $\sigma\in S_k$, its descent set is
$$Desc(\sigma):=\{i<k, \sigma(i)>\sigma(i+1)\}.$$
One can easily see that the action of $\mathcal D$ on $T(X)$ restricts to $T(X_n)$.

Let's go back to probability.
The first method of mixing that we have described is coded by the action of $\Psi_{elem}:=(p_1\ast p_{n-1})/n$ on $T_n(X_n)$: we extract the first card from the deck, then we insert it in all possible ways in the deck of remaining cards. The second method, that of standard mixings, is algebraically even more natural to describe: it is given by the action on $T_n(X_n)$ of $\Psi_{standard}:=(Id\ast Id)/2^n$. 
The problem of describing the Markov chain associated with this or that mixing method then comes down to the purely algebraic problem of computing the iterated composition of elements in $\mathcal{D}$  and, if we are interested in the asymptotic behavior of the chain, in computing their spectrum. 

Let us explain the case of the standard mixing, which has been studied in detail by Diaconis, Pang and Ram \cite{dpr}. Our proof is based on \cite{pat92,pat94}, we will see later that it applies to other asymptotic phenomena. Let us put $\Psi^k=Id^{\ast k}$, $J=Id-\nu$, $J_k:=J^{\ast k}$ and let us admit that we have verified that $\Psi^k\circ \Psi^l=\Psi^{kl}$ (the identity relies only on the commutativity of the product $\shuffle$ and on the bialgebra structure of $T_n(X)$). Recall also that the Stirling numbers of the first kind $s(j,i)$ are implicitly defined by $$x(x-1)\dots (x-j+1)=\sum\limits_{i=1}^js(j,i)x^i.$$
We obtain 
$$\Psi^k=(J+\nu)^{\ast k}=\sum\limits_{j=0}^k{k \choose j}J_j=\sum\limits_{j=0}^\infty{k \choose j}J_j$$
$$=\nu +\sum\limits_{j=1}^\infty\sum\limits_{i=1}^j [s(j,i)\frac{J_j}{j!}]k^i=:\sum\limits_{i=0}^\infty e^ik^i,$$
where $e^0=\nu$. From
$\Psi^k\circ \Psi^l=\Psi^{kl}$ we then deduce that $e^i$ is a spectral projector for $\Psi^k$ associated to the eigenvalue $k^i$. 

Let us mention that the first non trivial spectral projector $e^1$ can be used to solve the Baker-Campbell-Hausdorff problem. Indeed, up to the action of the set automorphisms $\sigma\longmapsto\sigma^{-1}$ of the symmetric groups, expanding $e^1$ as a linear combination of permutations, one recovers indeed the Mielnik-Pleba\'nski formula \cite{MPl} --- on approaches to the BCH problem using descents, the related notion of noncommutative symmetric functions and Hopf algebraic structures on the tensor algebra, see e.g. \cite{Reutenauer,Gelfand} and \cite[Remark 5.2.2]{CP21}. 

By specializing the calculation above to the vector space of linear combinations of words on the alphabet $X_n$ of length $n$ and without repetitions of the letters, and with some elementary arguments of linear algebra, we obtain that $\Psi_{standard}=\Psi^2/2^n$ has $1,1/2,...,1/2^n$ as eigenvalues. The spectral hole $1-1/2$ gives the speed of convergence to the equilibrium. This result on the spectrum of the $\Psi^k=Id^{\ast k}$ is not specific to $T(X_n)$ or $T(X)$: in technical terms, it holds for all connected graded bialgebras over a field of characteristic zero, commutative or cocommutative. This is in this group-theoretical context (commutative bialgebras are algebras of functions on groups and the operator $\Psi^k$ encodes in this context the ``dilation'' $x\longmapsto x^k$) that the ideas we have just described have actually been first developed, see \cite{pat93,pat94,CP21}. 

In the next section we will see that the same combinatorial arguments apply {\it mutatis mutandis} to a general form of the central limit theorem (CLT).

\subsection{Classical and quantum central limit theorems}

This part of the article follows very largely the ideas of von Waldenfels, Sch\"urmann, Speicher and the presentation thereof given by Lenczewski \cite{vWal,Schu,Spei1,Spei2,Lenc}.
 One can also find in these references an explanation of the links between the ``quantum'' central limit theorems (CLTs) and different forms of Fock spaces (the spaces on which the operators of annihilation and creation of particles in second quantization operate), links that we only mention here. Our impression is that these ideas deserve to be better known. We will present them at the light of our previous developments on algebraic and coalgebraic structures in classical and quantum probability.

We will focus here on the central limit theorem in its coalgebraic form, due to Sch\"urmann, which allows to generalize the theorem to many situations beyond its classical statement, but the general message is that there are some universal algebraic ideas and techniques that pop up again and again in different situations.
 We consider again
algebras of noncommutative random variables: associative algebras $\mathcal A$ with a state operator, i.e. a unitary linear form $\varphi$. 
We assume again that elements $a\in \mathcal A$ have moments $\varphi(a^n)$ at all orders.
The ``central limit problem'' consists then in studying the limit distribution of expressions of the type
$$S_n:=\frac{X_1+\dots X_n}{\sqrt n},$$
where the $X_i$ are independent and identically distributed random variables. The precise expression of the solution thus depends on the chosen notion of independence, but the general form of the solution does not: very much as the spectral properties of the ``dilation'' operators $\Psi^k$ hold for all connected graded commutative or cocommutative algebras (algebras of functions on prounipotent groups, and their duals), Sch\"urmann's CLT holds for all graded connected coalgebras. 

\subsubsection{The classical CLT}
In the classical case where the algebra $A$ is commutative and equipped with an expectation operator denoted $\mathbb E$ we know that, if the $X_i$ are centered, the limiting distribution of the $S_n$ will follow a centered Gaussian distribution, of variance the variance of the $X_i$.

So let us start by analyzing this case (the same calculation and arguments apply to tensor independence): we will find there almost exactly the algebraic structures appearing when studying cards sufflings. In its algebraic form, the (classical) independence of two subalgebras $A_1,A_2$ of $A$ is given by the property: $$\forall a_1\in A_1,\ a_2\in A_2, \ \ \mathbb E(a_1a_2)=\mathbb E(a_1)\mathbb E(a_2).$$
Given $(A,\mathbb{E})$, to create two independent copies of it, it is thus enough to consider the tensor product $A\otimes A$ equipped with the expectation $\mathbb{E}(a\otimes b):=\mathbb{E}(a)\mathbb{E}(b)$. The linear operator $$a\longmapsto {a\otimes 1+1\otimes a}$$ creates two independent copies of the random variable $a$ in $A\otimes A$. More generally,
$$a\longmapsto {a\otimes 1\otimes\dots\otimes 1+\dots +1\otimes\dots\otimes 1\otimes a}$$ creates $k$ copies of $a$ in $A^{\otimes k}$.

Let us abstract a bit more: consider $T(A)=\bigoplus\limits_{n=0}^\infty A^{\otimes n}$, the tensor algebra over $A$ and write the tensors $a_1 \otimes \dots a_n$ as words $a_1 \dots a_n$. In order not to confuse the tensor $a_1 \dots a_n$ with the product of the $a_i$ in $A$, we will write the latter $a_1\cdot_A\dots\cdot_A a_n$. We then equip $T(A)$ with the product of concatenation of words and the coproduct 
$$\Delta (a_1 \dots a_n)=\sum\limits_{I\coprod J=[n]}(a_I\otimes a_J),$$
where if $I=\{i_1,\dots,i_k\}$ with $i_j<i_{j+1}$, we set $a_I:=a_{i_1}\dots a_{i_k}.$
This coproduct is the dual of the shuffle product we met earlier, it provides $T(A)$ with a bialgebra structure. 
Its iterated action on an element $a$ of $A$ gives
$$\Delta_k(a)=a\otimes 1_{\mathbb C}\otimes 1_{\mathbb C}+\dots +1_{\mathbb C}\otimes 1_{\mathbb C}\otimes a.$$

The unitary linear form $\phi_T:a_1\dots a_n\longmapsto \mathbb{E}(a_1\cdot_A \dots \cdot_A a_n)$ (with thus $\phi_T(1_{\mathbb K})=1_{\mathbb K}$) makes $T(A)$ a noncommutative probability space. It extends to $T(A)\otimes T(A)$ (and by the same process to the higher tensor powers of $T(A)$) by :
$$\phi_T(w\otimes w')=\phi_T(w)\phi_T(w').$$ 

From the fact that $\phi_T$ is a unitary linear form, we deduce that we have, for random variables $a_1,\dots, a_k$,
$$\gamma_n(a_1,\dots,a_k):=\phi_T [\Delta_n(a_1)\Delta_n(a_2)\dots \Delta_n(a_k)]$$
$$=\sum\limits_{i=1}^k{n\choose i}\sum\limits_{\pi_1\coprod\dots\coprod\pi_i=[k]}\phi_T(a_{\pi_1})\dots \phi_T(a_{\pi_i}),$$
where $\pi_1\coprod\dots\coprod\pi_i=[k]$ denotes an ordered partition of $[k]$ whereas $\pi_1\cup\dots\cup\pi_i=[k]$ will denote an unordered partition.
Let us set $$\mu_i:=\frac{1}{i!}\sum\limits_{\pi_1\coprod\dots\coprod\pi_i=[k]}\phi_T(a_{\pi_1})\dots \phi_T(a_{\pi_i}).$$ We get the exact expansion (identical to the one obtained for bialgebras, excepted for replacing the operator $J_i/i!$ by the scalar $\mu_i$):
\begin{equation}
\gamma_n(a_1,\dots,a_k)=\sum\limits_{i=1}^k\sum\limits_{j=1}^is(i,j)n^j\mu_i=\sum\limits_{j=1}^k(\sum\limits_{i=j}^ks(i,j)\mu_i)n^j.
\end{equation}

When the variables $a_i$ are centered, $\phi_T(a_i)=\mathbb{E}(a_i)=0$, and partitions containing a singleton do not contribute to the sum, so that $\mu_i=0$ when $i>k/2$. Assume that $k=2p$ (the odd case is treated similarly).
The leading term  is then obtained when $i=j=p$, that is when all blocks are of cardinal 2, it is: 
$$s(p,p)n^p\mu_p=\mu_pn^p.$$
Let us call Wick partition of $[2p]$ the partitions $\pi=\pi_1\cup\dots\cup\pi_p$ whose blocks are all of cardinal 2, and let us then write $\pi\in Wick([2p])$; we obtain 
the Theorem 
\begin{theorem}[Classical CLT]
For centered random variables $a_1,\dots, a_{k}$,
$$\lim\limits_{n\to \infty}\frac{\phi_T}{\sqrt{n}^{k}}(\Delta_n(a_1)\Delta_n(a_2)\dots \Delta_n(a_{k}))=\sum\limits_{\pi\in Wick([k])}\phi_T(a_{\pi_1})\dots \phi_T(a_{\pi_{k/2}}),$$
if $k$ is even. The limit is zero when $k$ is odd.
\end{theorem}

When all $a_i$ are equal to a random variable $a$, we have obtained an algebraic formulation of the usual central limit theorem (valid also in the tensor independence case). In this case, $\frac{\phi_T}{\sqrt{n}^{k}}(\Delta_n(a_1)\Delta_n(a_2)\dots \Delta_n(a_{k}))$ indeed computes $\mathbb{E}(\frac{b_1+\dots +b_n}{\sqrt{n}})^k$, where the $b_i$ are independent copies of $a$, while, by Wick's theorem, the right term computes the $k$-th moment of a centered normal variable of variance the variance of $a$.

\subsubsection{The free CLT}

The transition from classical probability to the various quantum probability theories is not obvious in practice. It depends, as we said, first on the choice of a notion of independence, closely related to the corresponding notion of cumulants (on this point, see e.g. \cite{manzel}), as we already mentioned. We account briefly here for the example probably the most treated in the literature, that of free probabilities, introduced by D. Voiculescu \cite{voiculescu_92,voiculescu_95}.
\begin{defn}
Given a noncommutative probability space $(\mathcal A,\varphi)$, we say that the subalgebras $\mathcal A_1,\dots,\mathcal A_n$ are freely independent if and only if for all
$a_1,\dots,a_k$ in $\mathcal A_{i_1},\dots,\mathcal A_{i_k}$ with $i_j\not= i_{j+1}$, $j=1,\dots,k-1$ and $\varphi(a_i)=0$ for all $i$, we have
$$\varphi(a_1\cdot_{\mathcal A}\dots \cdot_{\mathcal A} a_n)=0.$$
\end{defn}

This assumption suffices to compute all $\varphi(b_1\cdot_{\mathcal A}\dots \cdot_{\mathcal A}  b_m)$ where each $b_i$ belongs to one of the subalgebras $A_1,\dots,A_n$ (without the centering assumption $\varphi(b_i)=0$) as soon as we know the value of $\varphi$ on the $A_i$.
For example, if $a$ and $b$ are in two freely independent subalgebras,
$$\varphi(abab)=\varphi(a)^2\varphi(bb)+\varphi(aa)^2-\varphi(a)^2\varphi(b)^2.$$
In this context, the central limit theorem, due to Speicher, is stated as follows. Although based in the end on a counting argument of partitions similar to the one we have given for the classical case, it also requires taking into account the form taken by the evaluation of $\varphi$ on arbitrary products of random variables belonging to independent algebras \cite{Spei2}.
We denote $Winc([2k])$ the non-crossing Wick partitions of $[2k]$ and we admit that we know how to create freely independent copies of an algebra of noncommutative random variables --- the procedure is an adaptation of the classical case where the tensor product of algebras has to be replaced by their free product.
\begin{theorem}
Let $a_1,\dots,a_k\in \mathcal A$ be centered random variables (that is, such that $\varphi(a_i)=0)$ ; $\mathcal A_1,\dots,\mathcal A_l,\dots$ be freely independent copies of $\mathcal A$. Let $a_i^j$ be a copy of $a_i$ in $\mathcal A_j$ and let
$$S_i^n:=\frac{a_i^1+\dots +a_i^n}{\sqrt{n}},$$
then 
$$\lim\limits_{n\to\infty}\varphi(S_1^n...S_k^n)=\sum\limits_{\pi\in Winc(k)}\varphi(a_{\pi_1})\dots \varphi(a_{\pi_{k/2}})$$
if $k$ is even and is equal to $0$ otherwise.
\end{theorem}

\subsubsection{A general coalgebraic CLT}
We will see later in this article how to associate bialgebras of sentences with free probability, which provides a way to directly connect free probability calculations with coalgebraic structures. Still another (and different) connection between the various quantum probability theories and coalgebras goes through the use of algebraic groups \cite{friedrich,manzel}.

We focus here instead  M. Sch\"urmann's central limit theorem for  coalgebras \cite{Schu}. It is a universal form of the central limit theorem and we find it interesting to include it in this survey as it is an example of how quantum probability ideas and techniques can potentially fertilize other fields, as the theorem could apply also to coalgebras different from those that appear in probability (although we are not aware of already existing applications of the theorem outside probability).

Let $C$ be a connected graded coalgebra, i.e. such that $C=\bigoplus\limits_{n=0}^\infty C_n$ with $C_0=\mathbb C$, the ground field, and $$\Delta: C_n\to \bigoplus\limits_{i+j=n} C_i\otimes C_j.$$

\begin{theorem}[Coalgebraic CLT]
We have, for any linear form $\varphi$, unitary and vanishing on $C_1$  and any $w\in C_k$ with $k$ even
$$\lim\limits_{n\to \infty}\varphi^{\ast n}(\frac{w}{\sqrt{n}^{k}})=\exp^\ast \kappa_\varphi(w),$$
where $\kappa_\varphi$ is the linear form on $C$ which cancels on $C_j$, $j\not=2$ and equals $\varphi$ on $C_2$.
\end{theorem}
We followed Lenczewski's formulation of the theorem \cite{Lenc}, Sch\"urmann's original formulation was more general, allowing $\varphi$ to vanish on $C_i$, $0<i<k$ --- the linear form $\kappa_\varphi$ cancels then on $C_j$, $j\not=k$ and the scaling factor should be $n^{-\frac{1}{k}}$ instead of $\sqrt{n}$.
The hypothesis that $\varphi$ vanishes on $C_1$ amounts, in probabilistic language, to consider only centered random variables.
The product $\ast$ used in both members of the equation is the convolution product of linear forms on a coalgebra.

Let us sketch how the theorem can be proved using the same computation as the one we performed to prove convergence to the uniform distribution in cards shufflings and in the classical case of the CLT (the following calculation till Eq. (\ref{hook}) is taken from \cite{CelP22}, where the reader can find various applications of the use of iterated coproducts in quantum probability).
Let $C$ be as above. We write $\nu$ for the projection onto $C_0$ orthogonally to the $C_i$, $i>0$, and write $Id=J+\nu$. The reduced coproduct on $C$ is defined by
$$\overline\Delta(c):=\Delta(c)-c\otimes 1-1\otimes c=J^{\otimes 2}\circ \Delta.$$
The last identity follows from the fact that $C$ is counital, this is the key ingredient of the following computations.
Similarly, the iterated coproduct and the iterated reduced coproduct (the iteration $k-1$ times of the reduced coproduct, mapping $C$ to $C^{\otimes k}$, written $\overline{\Delta}_k$) are related by the identity:
$$\overline{\Delta}_k=J^{\otimes k}\circ \Delta_k.$$
Now,
$$\Delta_k=(\nu+J)^{\otimes k}\circ \Delta_k$$
$$=\left(\sum\limits_{l< k}\sum\limits_{1\leq i_1<\dots<i_l\leq k}J\otimes\dots\otimes \nu\otimes\dots\otimes\nu\otimes\dots\otimes J\right)\circ \Delta_k,$$
where the term in the last summation formula contains $l$ copies of $\nu$, in positions $i_1,\dots,i_l$ (with possibly $i_1=1$ and/or $i_l=k$, that is, $\nu$ is allowed to be in first or last position in the tensor product).

Let now $f:[k]\hookrightarrow[n]$ be an increasing injection. We denote $\hat{f}$ the map from  $C^{\otimes k}$ to $C^{\otimes n}$ defined by 
$$\hat{f}(c_1\otimes\dots\otimes c_k):=d_1\otimes\dots\otimes d_n$$
with $d_{f(i)}:=c_i$ for $1\leq i\leq k$, and $d_j:=1$ if $j$ is not in the image of $f$.
Since $\Delta$ is counital, the above expression of $\Delta_{k}$ rewrites 
\begin{equation}\label{hook}
\sum\limits_{l< k}\sum\limits_{f:[k-l]\hookrightarrow[k]}\hat{f}\circ \overline{\Delta}_{k-l}
\end{equation}

Now, let $\phi$ be a unital linear form on $C$ (we don't require it to vanish on $C_1$). Then, with the same notation, as $\phi$ is unital, $$\phi^{\otimes k}\circ \hat{f}\circ \overline{\Delta}_{k-l}=\phi^{\otimes k-l}\circ \overline{\Delta}_{k-l}.$$
Besides, given $w\in C_p$, for degree reasons we have $\overline{\Delta}_{l}(w)=0$ for $l>p$. By definition of the convolution product, we get
$$\phi^{\ast k}(w)=\phi^{\otimes k}\circ \Delta_k(w)=\phi^{\otimes k}(\sum\limits_{l\leq k}\sum\limits_{f:[k-l]\hookrightarrow[k]}\hat{f}\circ \overline{\Delta}_{k-l}(w))$$
$$=\sum\limits_{i=1}^p{k\choose i}\phi^{\otimes i}\circ\overline{\Delta}_{i}(w).$$
Applying our previous reasoning to rewrite the binomial coefficients using Stirling numbers of the first kind allows to recover the coalgebraic CLT.

\newpage

\section{From words to phrases}\label{phrases}
We have seen how the combinatorics of words can be extended from Lie algebra to preLie algebra computations and how the properties of the spectrum of dilations (the $\Psi^k$ operators) can be extended from the tensor algebra (and other cocommutative connected bialgebras) to the study of phenomena such as the (exact, non asymptotic) study of central limit theorems in a very general setting that encompasses the central limit theorem for coalgebras.

We come back in this section on the role of cumulants in quantum probabilities, and then explain how the noncommutative universe invites the definition of new bialgebra structures where words are replaced by sentences (sequences of words) as their buiding blocks.
Orthogonal polynomials, another classical subject in algebraic combinatorics, will serve as a guideline, as well as Wick polynomials. 

Let us mention that one of the reasons for chosing Wick polynomials as an illustration is driven by the general idea of revisiting algebraically fundational ingredients of quantum mechanics: the seminal work of Wick was indeed driven, besides Wick's interest for abstract results and structures, by the necessity to perform from scratch renormalization in quantum field theory through the introduction of normal products (``removing the infinite charge of the negative sea", in his own terms).
Our presentation is largely inspired by two articles of the second author with K. Ebrahimi-Fard, N. Tapia and L. Zambotti \cite{eptz0,eptz}.

\subsection{Wick polynomials from bialgebras}
Let us first recall how cumulants relate to Hermite polynomials. 
Hermite polynomials are defined by physicists by the formula
   $$ H_{n}(x)=(-1)^{n}e^{x^{2}}{\frac {d^{n}}{dx^{n}}}e^{-x^{2}}.$$
  They appear naturally when solving the stationary Schrödinger equation for a harmonic potential in $L^2(\mathbb{R})$ and are intimately linked to the notion of free quantum system in second quantization \cite{Zeidler}.
Probabilists define them as
$$ {He}_{n}(x)=(-1)^{n}e^{\frac {x^{2}}{2}}{\frac {d^{n}}{dx^{n}}}e^{-{\frac {x^{2}}{2}}}.$$
When $\Omega$ is a normal random variable, the $ {He}_{n}(\Omega)$ define a Hilbertian basis of $L^2(\Omega)$. One speaks of ``chaos decomposition'' for the associated orthogonal decomposition of $L^2(\Omega)$ \cite{peccati}.
    We pass from one definition to the other by 
$$H_{n}(x)=(\sqrt{2})^n{\mathit {He}}_{n}\left({\sqrt {2}}\,x\right).$$
In the following, we will focus on the probabilistic point of view, but we want to underline the general principle, valid for most of the developments of this chapter, of the possibility of a translation from one language (quantum physics) into the other (probabilities, classical or non commutative).

Hermite polynomials are a special case of Wick polynomials. The Wick polynomial $W_n(x)$ associated to any random variable $X$ (provided it has moments at all orders) is defined by $W_0(x)=1$,
the condition that defines Appell polynomials:
 $$ \frac{\mathrm d}{\mathrm dx}W_n(x)=nW_{n-1}(x),$$
and $\mathbb E( W_n(X))=0$.
for all $n>0$. When $X=\Omega$, a standard normal random variable (centered, with variance 1), we have: $He_n(x)=W_n(x)$.

Recall that $K(t) := \sum_{n>0} c_n \frac{t^n}{n!}$,
the exponential generating series  of cumulants associated to $X$,
 is defined by the identity
$M(t)=exp ( K(t) ),$ where $M(t)$ stands for the exponential generating series of momenta.
The link between Wick polynomials and cumulants is given by the identity
$$\sum_{n\ge0}W_n(x)\frac{t^n}{n!}
  =\frac{\exp({tx})}{\mathbb E (\exp({tX}))}=\exp({tx-K(t)}).$$

These constructions can be interpreted using the language of bialgebras, which also allows to extend them to a very general setting as it encompasses the study of arbitrary unital linear forms on (graded) bialgebras. In this section we denote $\mathbb K$ an arbitrary ground field.

Let now $(H=\bigoplus\limits_{n\in\mathbb N}H_n,\cdot,\Delta)$ be an arbitrary graded bialgebra and $\varphi$ an arbitrary linear form on $H$ that we assume unital ($\varphi(1)=1$) and convolution invertible. 

The standard situation is the case where $H_0=\mathbb K$: any unital linear form on $H$ is then convolution invertible. The inverse is then given by $\varphi^{*-1}=\frac{\nu}{\nu+(\varphi-\nu)}=\sum\limits_{n=1}^\infty(-1)^n(\varphi-\nu)^{\ast n}$. 
We used the usual notation:  $\nu$ stands for the canonical projection onto $H_0$. For degree reasons, the infinite sum reduces to a finite sum when it is applied on an element of $H$, so that the inverse is well-defined. We used the notation $\varphi^{*-1}$ for the {\it convolution} inverse, not to be confused with a composition inverse of $\varphi$ (that never exists if $H$ is different from the ground field!). 

The convolution product $W:=\varphi^{*-1}\ast Id$ defines then a linear automorphism of $H$. This follows from the identity:
$$(\varphi\ast Id)\circ (\varphi^{*-1}\ast Id)=\varphi^{*-1}\ast \varphi\ast Id=Id.$$
By structure transportation, this automorphism can be used to deform the product and the coproduct on $H$. For example, a new associative product $\cdot_W$ on $H$ is defined by
$W(x)\cdot_WW(y):=W(x\cdot y),$ or $$x\cdot_W y:=W(W^{-1}(x)\cdot W^{-1}(y)).$$
These products generalize the Wick product of Gaussian variables and quantum fields (called also normal ordered product in physics).

The case of Hermite polynomials is obtained as follows (see \cite{eptz0} for details and a proof). Consider the (canonically graded) algebra of polynomials $\mathbb R[x]$ with the coproduct $\Delta(x^n)=\sum\limits_{k=0}^n{n\choose k}x^{k}\otimes x^{n-k}$. Set $\varphi(x^n):=\mathbb E(\Omega^n)$. Then, one can show that $$W(x^n)=W_n(x),$$
where the left term is defined as $\varphi^{*-1}\ast Id(x^n)$ and the right term is the $n$-th Hermite polynomial. 
Wick polynomials for non Gaussian variables can be recovered by the same process. 

\subsection{Quantum Wick polynomials}

The construction we have described in the previous section leads immediately to the generalization of Wick polynomials to the noncommutative framework of tensor independence in quantum probability (see \cite{eptz}). Formulas and constructions are then very similar to the classical case: we will therefore focus instead on the other quantum probability theories that require the introduction of new ideas.

We will limit ourselves here to presenting a result: the definition of noncommutative polynomials which are associated to random variables in free probability in the same way as Hermite polynomials are to Gaussian variables and, more generally, more general Wick polynomials to classical random variables. The idea goes back to the work of M. Anshelevich on the non-commutative Appell polynomials \cite{A1,A2,A3}, but we will follow the algebraic approach of \cite{eptz}, where we will see the reappearance of various ideas we have already encountered: bialgebra structures, word combinatorics, shuffles... which can be used systematically to study many other algebraic problems encountered in the context of noncommutative probabilities.

Given $(\mathcal A,\varphi)$ an algebra of noncommutative random variables, we denote $T^+(\mathcal A)$ the nonunitary tensor algebra on $\mathcal A$ :
\[
  T^+(\mathcal A):=\bigoplus_{n>0} \mathcal A^{\otimes n}.
\]
Its elements are written as usual as linear combinations of words, $a_1\cdots a_n:=a_1\otimes\dots \otimes a_n$. Let us then introduce the double tensor algebra, $T(T^+(\mathcal A))$, over $\mathcal A$ : 
\[
  T(T^+(\mathcal A)):=\bigoplus_{n\geq 0} (T^+(\mathcal A))^{\otimes n}.
\]
Its elements are linear combinations of tensors $W=w_1\otimes \dots\otimes w_n$, where the $w_i$ are elements of $T^+(\mathcal A)$ of the form $a_1\cdots a_n$. We represent $W$ as a sentence $w_1|w_2|...|w_n$ using the bar notation of topologists to distinguish the two levels of tensor products (for example $(a\otimes b)\otimes (c\otimes d)\in (\mathcal A^{\otimes 2})\otimes (\mathcal A^{\otimes 2})$ is written $w_1|w_2$ with $w_1=ab,\ w_2=cd$). The product in $T(T^+(\mathcal A))$ is the concatenation of sentences: $w_1|w_2|...|w_n\cdot_{T(T^+(\mathcal A))}w_1'|w_2'|...|w_m'=w_1|w_2|...|w_n|w_1'|w_2'|...|w_m'.$
We transform $T(T^+(\mathcal A))$ into an algebra of noncommutative random variables by defining $\Phi \colon T(T^+(\mathcal A))\to \mathbb{C}$ as the unique unitary and multiplicative extension of the linear form $\varphi$ defined on $T^+(\mathcal A)$ by $\varphi(a_1\dots a_n) := \varphi(a_1\cdot_{\! \scriptscriptstyle{\mathcal A}} \dots\cdot_{\!\scriptscriptstyle{\mathcal A}} a_n)$ :
$$\Phi(w_1|w_2|...|w_n):=\varphi(w_1)\dots \varphi(w_n).$$

Now let $U\subset\mathbb N^\ast$. A ``connected component'' of $U$ is a maximal sequence of consecutive integers in $U$. For $S\subseteq[n]$, let $J_1^S,\dotsc,J_k^S$ be the connected components of $[n]\setminus S$, ordered in the natural order of their minimal elements.

\begin{defn}\label{def:coprod}
We then define $\Delta\colon T^+(\mathcal A)\to {T}(\mathcal A)\otimes T(T^+(\mathcal A))$ by
\begin{equation}
\label{eq:theCoprod}
  \Delta(a_1\cdots a_n)
  := a_1\cdots a_n \otimes 1 
  			+ 1 \otimes a_1\cdots a_n 
	+ \sum_{\substack{S\subsetneq[n] \neq \emptyset}}
				a_S\otimes a_{J^S_1}\vert\cdots\vert a_{J^S_k}.
\end{equation}
\end{defn}

This coproduct is very close to the dual coproduct to the shuffle product in the tensor algebra. If we think of a deck of cards, we let the reader convince himself that this coproduct amounts  to doing the following operation: given a deck of cards, recursively take the top card and put it randomly in a deck on the right or on the left {\it but} creating a new deck on the right each time we just put a card on the left.

This application extends uniquely into an application $\Delta\colon T(T^+(\mathcal A))\to T(T^+(\mathcal A))\otimes T(T^+(\mathcal A))$ if we require that $\Delta(1)=1\otimes 1$ and that $\Delta(w_1|w_2|...|w_n):=\Delta(w_1)\dots \Delta(w_n).$
One can show that the algebra $T(T^+(\mathcal A))$ with the coproduct $\Delta$ is a bigebra (neither commutative nor cocommutative).
As $\Phi$ is a unital linear form, it is an invertible element in the convolution algebra of linear endomorphisms of $T(T^+(\mathcal A))$. Its inverse for the convolution product is denoted $\Phi^{*-1}$. 

\begin{defn}
  \label{dfn:fWick}
  The free Wick morphism $\mathrm{W} \colon T(T^+(\mathcal A))\to T(T^+(\mathcal A))$ is defined by
  \[
    \mathrm{W} :=({Id}\otimes\Phi^{*-1})\Delta.
  \]
\end{defn}
We call \emph{free Wick polynomials} (free Appell polynomials in the terminology of M. Anshelevich)
the $\mathrm{W}(a_1\cdots a_n)$, $a_i\in \mathcal A$, $i=1,\ldots, n$.

From this definition, we deduce that the words of $T^+(\mathcal A)$ can re-expressed in terms of the free Wick polynomials:
\begin{equation}
  a_1\dotsm a_n=\sum_{S\subseteq[n]}  \mathrm{W}(a_S)\Phi(a_{J_1^S}|\dots |a_{J_k^S}).
\end{equation}

We find, in the lower degrees:
\begin{align}
  \mathrm{W}(a_1) &= a_1-\varphi(a_1) , \nonumber\\
  \mathrm{W}(a_1a_2) &= a_1a_2-\varphi(a_2)a_1-\varphi(a_1)a_2
        - \big(\varphi(a_1\cdot_{\!\scriptscriptstyle{A}} a_2)-2\varphi(a_1)\varphi(a_2)\big),  \nonumber\\
  \begin{split}
  \mathrm{W}(a_1a_2a_3) &= a_1a_2a_3-\varphi(a_3)a_1a_2-\varphi(a_2)a_1a_3-\varphi(a_1)a_2a_3\\
  &\,\, - \big(\varphi(a_2\cdot_{\!\scriptscriptstyle{A}} a_3)
    - 2\varphi(a_2)\varphi(a_3)\big)a_1
      + \varphi(a_1)\varphi(a_3)a_2
    - \big(\varphi(a_1\cdot_{\!\scriptscriptstyle{A}} a_2)\\
  &\,\, -2\varphi(a_1)\varphi(a_2)\big)a_3
    - \big(\varphi(a_1\cdot_{\!\scriptscriptstyle{A}} a_2\cdot_{\!\scriptscriptstyle{A}} a_3)
    - 2\varphi(a_1)\varphi(a_2\cdot_{\!\scriptscriptstyle{A}} a_3)\\
  &\,\, -2\varphi(a_3)\varphi(a_1\cdot_{\!\scriptscriptstyle{A}} a_2)
    - \varphi(a_2)\varphi(a_1\cdot_{\!\scriptscriptstyle{A}} a_3)
  + 5\varphi(a_1)\varphi(a_2)\varphi(a_3)\big).
  \end{split} 
\end{align}

The free Wick polynomial $\mathrm{W}(a_1\dotsm a_n)$, can also be rewritten in terms of the free cumulants \cite{A1,eptz0}
\[
  \mathrm{W}(a_1\dotsm a_n)=\sum_{S\subseteq[n]}a_S\sum_{\substack{\pi\in\operatorname{I}
  ([n]\setminus S)\\\pi\cup S\in \operatorname{NC}([n])}}(-1)^{|\pi|}\prod_{B\in\pi}\kappa(a_B).
\]
We can see in the articles \cite{A1,A2,A3,eptz0} that the theory of free Wick polynomials has innumerable properties and is very natural, both from the point of view of the theory of polynomials in noncommutative variables and from the point of view of the use of coalgebras and bialgebras in noncommutative probabilities. 

As Wick calculus has played an essential role in quantum field theory and in classical probability since Wick's seminal 1950 paper \cite{Wick},
it is quite intellectually satisfying to see that it also carries over naturally to the context of quantum probabilities. We have developed this example because it is less well understood than other aspects of noncommutative probabilities: there are probably still aspects of the noncommutative Wick calculus to be explored.


\newpage

\section{Insufficiency of classical probabilities : A tribute from Bell to Aspect. }
\label{sec:PQVPC}

In this final section, we will come back on  a historical argument that leads most physicists  to conclude that classical probabilities are insufficient, or in other words, that classical probabilities are \textbf{strictly included} in quantum probabilities.
We will follow Bell  \cite{bell}, who was the first to try to establish quantitative arguments to settle the question. 

Before this, we start by presenting the iterative measurement of a qubit. We will see that it leads to a first notch at the quest for an understanding of quantum by classical probabilities.

\subsection{Two level quantum systems}

We consider a quantum system whose Hilbert space
$\mathcal{H}$ is bidimensional, with basis $\left(\left|+1\right\rangle ,\left|-1\right\rangle)\right)$.
Such systems model the smallest unit for quantum information. Physically, it models the spin $\frac{1}{2}$
of the electron or neutron, or the polarization of the photon. In the following, we use 
the language of the  spin $\frac{1}{2}$ rather than that of the photon polarization.

\begin{itemize}
\item We first introduce the spin measurement observables along the directions of space in a given reference frame: 

\[
\sigma_{X}\equiv\left(\begin{array}{cc}
0 & 1\\
1 & 0
\end{array}\right);\sigma_{Y}\equiv\left(\begin{array}{cc}
0 & -i\\
i & 0
\end{array}\right);\sigma_{Z}\equiv\left(\begin{array}{cc}
1 & 0\\
0 & -1
\end{array}\right).
\]

Similarly, the spin measurement observable according to the
general direction $\overrightarrow{n}\in R^{3}$ ($\overrightarrow{n}$
of norm unity : $\left\Vert \overrightarrow{n}\right\Vert =1$ ) is given by
by the matrix

\[
\sigma_{\overrightarrow{n}}\equiv\overrightarrow{n}.\overrightarrow{\sigma}=\left(\begin{array}{cc}
n_{3} & -in_{2}+n_{1}\\
in_{2}+n_{1} & -n_{3}
\end{array}\right).
\]
 
By representing $\overrightarrow{n}$ in spherical coordinates 
 $\overrightarrow{n}=\left(\begin{array}{c}
\sin\left(\theta\right)\cos\left(\varphi\right)\\
\sin\left(\theta\right)\sin\left(\varphi\right)\\
\cos\left(\theta\right)
\end{array}\right)$
 with 
  $0\leq\theta\leq\pi$ et $0\leq\varphi\leq2\pi$, and
 noting $\sigma_{\overrightarrow{n}}\equiv\sigma_{\theta,\varphi}$,
the spectral decomposition of this matrix is then given by 

\begin{equation}
\sigma_{\theta,\varphi}=\psi_{\theta,\varphi}\psi_{\theta,\varphi}^{+}-\omega_{\theta,\varphi}\omega_{\theta,\varphi}^{+}\textrm{ with }\begin{cases}
\psi_{\theta,\varphi}=\left(\begin{array}{c}
\cos\left(\frac{\theta}{2}\right)\exp\left(-i\varphi\right)\\
\sin\left(\frac{\theta}{2}\right)
\end{array}\right)\\
\omega_{\theta,\varphi}=\left(\begin{array}{c}
\sin\left(\frac{\theta}{2}\right)\exp\left(-i\varphi\right)\\
-\cos\left(\frac{\theta}{2}\right)
\end{array}\right)
\end{cases}.\label{eq:ePqubit}
\end{equation}

As we saw in the previous paragraph, since the spectrum of
$\sigma_{\theta,\varphi}$ is given by $\pm1$,  the ``outcomes
of the measure of $\sigma_{\theta,\varphi}$ are therefore bivalent  $\pm1$. 
This is the reason for the choice of the term ``qubit'', with the novelty compared to classical bit
that one can ask an infinity of different Yes/No questions
(indexed by $\theta;\varphi$) to a given qubit !
 \vspace{0.5cm}

\item The general state of a qubit is then given by the density matrix : 
\begin{equation}
\rho=\frac{1}{2}\left(\begin{array}{cc}
1+z & x-iy\\
x+iy & 1-z
\end{array}\right),\label{eq:bl-1}
\end{equation}
whose positivity is equivalent to the condition $x^{2}+y^{2}+z^{2}\leq1$.
The states of a qubit are thus in bijection with the unit ball of $\mathbb{R}^{3},$
called in this context the {\it Bloch ball}.

The pure states of the qubit $\rho=\psi\psi^{\dagger},$ (where $\psi = \alpha \left|+1\right\rangle+\beta \left|-1\right\rangle$ with $|\alpha|^2+|\beta|^2=1$)
belong then to the sphere of radius 1. They are represented in
spherical coordinates
$\left(\begin{array}{c}
x\\
y\\
z
\end{array}\right)=\left(\begin{array}{c}
\sin\left(\theta'\right)\cos\left(\varphi'\right)\\
\sin\left(\theta'\right)\sin\left(\varphi'\right)\\
\cos\left(\theta'\right)
\end{array}\right)$ 
(with $0\leq\theta'\leq\pi$ and $0\leq\varphi'\leq2\pi$),  by 
vectors $\psi_{\theta',\varphi'}$ given in the basis $(\left|+1\right\rangle , \left|-1\right\rangle)$ by (\ref{eq:ePqubit}).

\vspace{0.5cm}

\item In this pure state $\psi_{\theta',\varphi'}$, the measure of the spin according to
direction $\left(\theta,\varphi\right)$ is thus given by the
probability (\ref{eq:PEQ-1-2-1})
{\small
\begin{equation}
\mathbb{P}_{\psi_{\theta',\varphi'}}\left(\sigma_{\theta,\varphi}=1\right)=\left|\psi_{\theta',\varphi'}^{+}\psi_{\theta,\varphi}\right|^{2}=\left|\cos\left(\frac{\theta'}{2}\right)\cos\left(\frac{\theta}{2}\right)\exp\left(i\left(\varphi'-\varphi\right)\right)+\sin\left(\frac{\theta'}{2}\right)\sin\left(\frac{\theta}{2}\right)\right|^{2}.\label{eq:PEQ-1-1-1}
\end{equation}}

Moreover $\mathbb{P}\left(\sigma_{\theta,\varphi}=-1\right)=1-\mathbb{P}\left(\sigma_{\theta,\varphi}=1\right).$
Similarly, in the pure state $\omega_{\theta',\varphi'}$, the measurement
of the spin along the direction $\left(\theta,\varphi\right)$ is
given by the probability (\ref{eq:PEQ-1-2-1})
{\small
\begin{equation}
\mathbb{P}_{\omega_{\theta',\varphi'}}\left(\sigma_{\theta,\varphi}=1\right)=\left|\omega_{\theta',\varphi'}^{+}\psi_{\theta,\varphi}\right|^{2}=\left|\sin\left(\frac{\theta'}{2}\right)\cos\left(\frac{\theta}{2}\right)\exp\left(i\left(\varphi'-\varphi\right)\right)-\cos\left(\frac{\theta'}{2}\right)\sin\left(\frac{\theta}{2}\right)\right|^{2}.\label{eq:PEQ-1-1-1-1}
\end{equation}}

For example, if the initial state is the pure state $\psi_{0,0}=\left|+1\right\rangle $
and that we measure successively the spin $\sigma_{\theta_{1},0}$ and then
instantaneously after the spin $\sigma_{\theta_{2},0}$, 
taking into account the phenomenon of projection of the state on the ket $\psi_{\theta_{1},0}$ or
 $\omega_{\theta_{1},0}$ after the first measurement,
we obtain the probabilities
{\small
\begin{equation}
\begin{cases}
\mathbb{P}_{\left|+1\right\rangle }\left(\sigma_{\theta_{1},0}=1,\sigma_{\theta_{2},0}=1\right)=\mathbb{P}_{\left|+1\right\rangle }\left(\sigma_{\theta_{1},0}=1\right)\mathbb{P}_{\psi_{\theta_{1},0}}\left(\sigma_{\theta_{2},0}=1\right)=\cos^{2}\left(\frac{\theta_{1}}{2}\right)\cos^{2}\left(\frac{\theta_{2}-\theta_{1}}{2}\right)\\
\mathbb{P}_{\left|+1\right\rangle }\left(\sigma_{\theta_{1},0}=1,\sigma_{\theta_{2},0}=-1\right)=\mathbb{P}_{\left|+1\right\rangle }\left(\sigma_{\theta_{1},0}=1\right)\mathbb{P}_{\psi_{\theta_{1},0}}\left(\sigma_{\theta_{2},0}=-1\right)=\cos^{2}\left(\frac{\theta_{1}}{2}\right)\sin^{2}\left(\frac{\theta_{2}-\theta_{1}}{2}\right)\\
\mathbb{P}_{\left|+1\right\rangle }\left(\sigma_{\theta_{1},0}=-1,\sigma_{\theta_{2},0}=1\right)=\mathbb{P}_{\left|+1\right\rangle }\left(\sigma_{\theta_{1},0}=-1\right)\mathbb{P}_{\omega_{\theta_{1},0}}\left(\sigma_{\theta_{2},0}=1\right)=\sin^{2}\left(\frac{\theta_{1}}{2}\right)\sin^{2}\left(\frac{\theta_{2}-\theta_{1}}{2}\right)\\
\mathbb{P}_{\left|+1\right\rangle }\left(\sigma_{\theta_{1},0}=-1,\sigma_{\theta_{2},0}=-1\right)=\mathbb{P}_{\left|+1\right\rangle }\left(\sigma_{\theta_{1},0}=-1\right)\mathbb{P}_{\omega_{\theta_{1},0}}\left(\sigma_{\theta_{2},0}=-1\right)=\sin^{2}\left(\frac{\theta_{1}}{2}\right)\cos^{2}\left(\frac{\theta_{2}-\theta_{1}}{2}\right)
\end{cases}.\label{eq:MIQ}
\end{equation}
}

These calculations illustrate the non commutativity of the
 measurement seen in the previous paragraph (\ref{eq:M1},\ref{eq:M2}),
because for example
{\small
\[
\mathbb{P}_{\left|+1\right\rangle }\left(\sigma_{\theta_{1},0}=1,\sigma_{\theta_{2},0}=1\right)=\cos^{2}\left(\frac{\theta_{1}}{2}\right)\cos^{2}\left(\frac{\theta_{2}-\theta_{1}}{2}\right)\]
\[\neq\cos^{2}\left(\frac{\theta_{2}}{2}\right)\cos^{2}\left(\frac{\theta_{2}-\theta_{1}}{2}\right)=\mathbb{P}_{\left|+1\right\rangle }\left(\sigma_{\theta_{2},0}=1,\sigma_{\theta_{1},0}=1\right).
\]
}

\end{itemize}


If we add a third measure $\sigma_{\theta_{3},0}$ to the formulas
(\ref{eq:MIQ}), we find by using the formulas (\ref{eq:PEQ-1-1-1},\ref{eq:PEQ-1-1-1})
and taking into account the projection of the state after the first
and after the second measurement, according to their result, the probabilities : 

\textbf{\small{
\begin{equation}
\begin{cases}
\mathbb{P}_{\left|+1\right\rangle }\left(\sigma_{\theta_{1},0}=1,\sigma_{\theta_{2},0}=1,\sigma_{\theta_{3},0}=1\right)\\
=\mathbb{P}_{\left|+1\right\rangle }\left(\sigma_{\theta_{1},0}=1\right)\mathbb{P}_{\psi_{\theta_{1},0}}\left(\sigma_{\theta_{2},0}=1\right)\mathbb{P}_{\psi_{\theta_{2},0}}\left(\sigma_{\theta_{3},0}=1\right)=\cos^{2}\left(\frac{\theta_{1}}{2}\right)\cos^{2}\left(\frac{\theta_{2}-\theta_{1}}{2}\right)\cos^{2}\left(\frac{\theta_{3}-\theta_{2}}{2}\right),\\
\mathbb{P}_{\left|+1\right\rangle }\left(\sigma_{\theta_{1},0}=1,\sigma_{\theta_{2},0}=1,\sigma_{\theta_{3},0}=-1\right)\\
=\mathbb{P}_{\left|+1\right\rangle }\left(\sigma_{\theta_{1},0}=1\right)\mathbb{P}_{\psi_{\theta_{1},0}}\left(\sigma_{\theta_{2},0}=1\right)\mathbb{P}_{\psi_{\theta_{2},0}}\left(\sigma_{\theta_{3},0}=-1\right)=\cos^{2}\left(\frac{\theta_{1}}{2}\right)\cos^{2}\left(\frac{\theta_{2}-\theta_{1}}{2}\right)\sin^{2}\left(\frac{\theta_{3}-\theta_{2}}{2}\right),\\
\mathbb{P}_{\left|+1\right\rangle }\left(\sigma_{\theta_{1},0}=1,\sigma_{\theta_{2},0}=-1,\sigma_{\theta_{3},0}=1\right)\\
=\mathbb{P}_{\left|+1\right\rangle }\left(\sigma_{\theta_{1},0}=1\right)\mathbb{P}_{\psi_{\theta_{1},0}}\left(\sigma_{\theta_{2},0}=-1\right)\mathbb{P}_{\omega_{\theta_{2},0}}\left(\sigma_{\theta_{3},0}=1\right)=\cos^{2}\left(\frac{\theta_{1}}{2}\right)\sin^{2}\left(\frac{\theta_{2}-\theta_{1}}{2}\right)\sin^{2}\left(\frac{\theta_{3}-\theta_{2}}{2}\right),\\
\mathbb{P}_{\left|+1\right\rangle }\left(\sigma_{\theta_{1},0}=1,\sigma_{\theta_{2},0}=-1,\sigma_{\theta_{3},0}=-1\right)\\
=\mathbb{P}_{\left|+1\right\rangle }\left(\sigma_{\theta_{1},0}=1\right)\mathbb{P}_{\psi_{\theta_{1},0}}\left(\sigma_{\theta_{2},0}=-1\right)\mathbb{P}_{\omega_{\theta_{2},0}}\left(\sigma_{\theta_{3},0}=-1\right)=\cos^{2}\left(\frac{\theta_{1}}{2}\right)\sin^{2}\left(\frac{\theta_{2}-\theta_{1}}{2}\right)\cos^{2}\left(\frac{\theta_{3}-\theta_{2}}{2}\right).
\end{cases}\label{eq:MIQ-1}
\end{equation}}}

And the same for the probabilities starting with $\sigma_{\theta_{1},0}=-1$.
From these formulas, we can see that

{\small 
\begin{align*}
\mathbb{P}_{\left|+1\right\rangle }\left(\sigma_{\theta_{1},0}=1,\sigma_{\theta_{2},0}=1,\sigma_{\theta_{3},0}=1\right)+\mathbb{P}_{\left|+1\right\rangle }\left(\sigma_{\theta_{1},0}=1,\sigma_{\theta_{2},0}=-1,\sigma_{\theta_{3},0}=1\right)\\
=\cos^{2}\left(\frac{\theta_{1}}{2}\right)\cos^{2}\left(\frac{\theta_{2}-\theta_{1}}{2}\right)\cos^{2}\left(\frac{\theta_{3}-\theta_{2}}{2}\right)+\cos^{2}\left(\frac{\theta_{1}}{2}\right)\sin^{2}\left(\frac{\theta_{2}-\theta_{1}}{2}\right)\sin^{2}\left(\frac{\theta_{3}-\theta_{2}}{2}\right).
\end{align*}}
whereas the formula (\ref{eq:MIQ}) gives directly that 
\[
\mathbb{P}_{\left|+1\right\rangle }\left(\sigma_{\theta_{1},0}=1,\sigma_{\theta_{3},0}=1\right)=\cos^{2}\left(\frac{\theta_{1}}{2}\right)\cos^{2}\left(\frac{\theta_{3}-\theta_{1}}{2}\right).
\]
We have the bad surprise to note that 
\[
\mathbb{P}_{\left|+1\right\rangle }\left(\sigma_{\theta_{1},0}=1,\sigma_{\theta_{2},0}=1,\sigma_{\theta_{3},0}=1\right)+\mathbb{P}_{\left|+1\right\rangle }\left(\sigma_{\theta_{1},0}=1,\sigma_{\theta_{2},0}=-1,\sigma_{\theta_{3},0}=1\right)\]
\[\neq\mathbb{P}_{\left|+1\right\rangle }\left(\sigma_{\theta_{1},0}=1,\sigma_{\theta_{3},0}=1\right),
\]
that is to say that the rule of total probabilities, to which we are used since childhood, is false here!

After reflection, the problem comes from a remnant of
formula (\ref{eq:projC-1-1-1}) : making a measurement without reading the
result is not the same as not making a measurement at all ! 
The non validity of the total probability rule seems to be bad news for our search for a
 classical probabilistic interpretation of a quantum system.
However, this first experiment is insufficient
to totally condemn this quest. In particular because the measurements take place at the same place in space: nothing prevents therefore the second random variable (translating the second spin measurement)  from depending on (i.e. being influenced by) the angle chosen
in the first measurement, so $\sigma_{\theta_{2},0}(\theta_{1}),$
and so on for the third measurement which could depend on the first two
angles\label{subsec:arg1}. 
With this hypothesis, the previous reasoning is invalidated because we find
by the same arguments as above
\begin{align*}
\mathbb{P}_{\left|+1\right\rangle }\left(\sigma_{\theta_{1},0}=1,\sigma_{\theta_{2},0}\left(\theta_{1}\right)=1,\sigma_{\theta_{3},0}\left(\theta_{1},\theta_{2}\right)=1\right)\\
+\mathbb{P}_{\left|+1\right\rangle }\left(\sigma_{\theta_{1},0}=1,\sigma_{\theta_{2},0}\left(\theta_{1}\right)=-1,\sigma_{\theta_{3},0}\left(\theta_{1},\theta_{2}\right)=1\right),\\
\neq\mathbb{P}_{\left|+1\right\rangle }\left(\sigma_{\theta_{1},0}=1,\sigma_{\theta_{3},0}\left(\theta_{1}\right)=1\right),
\end{align*}
which, this time, does not break the total probability rule
since the events $\sigma_{\theta_{3},0}\left(\theta_{1},\theta_{2}\right)=1$
and $\sigma_{\theta_{3},0}\left(\theta_{1}\right)=1$ are different --in the first case an intermediate measurement was performed.

In order to overcome this problem, one idea is to ``delocalize'' the measurements
in 2 very distant points of space. This is what will be presented
below with the notion of Bell's pair of spins.

\subsection{Pure Bell pair state}
\subsubsection{Pure Bell state for a pair of spins $\frac{1}{2}$}
For a pair of spins $\frac{1}{2}$,
the Hilbert space of the pair is  $\mathcal{H}=\mathcal{H}_{A}\otimes\mathcal{H}_{B}$
: the tensor product of the Hilbert space  $\mathcal{H}_{A}$ bidimensional with basis $\left(\left|+1\right\rangle _{A},\left|-1\right\rangle _{A}\right),$
with the Hilbert space $\mathcal{H}_{B}$ bidimensional with basis$\left(\left|+1\right\rangle _{B},\left|-1\right\rangle _{B}\right).$
In the following, for the sake of simplicity, we will not note the indices $A$ and $B$.  

We then define a particular pure state, the pure state of a Bell spin pair, which is
given by the vector 

\begin{equation}
\label{etatpur}
\left|Bell\right\rangle \equiv\frac{\left|-1\right\rangle \otimes\left|1\right\rangle -\left|1\right\rangle \otimes\left|-1\right\rangle }{\sqrt{2}}.
\end{equation}
It is obvious that the pure state $\left|Bell\right\rangle $
cannot be written as a tensor product of 2 pure states (one in $\mathcal{H}_{A}$, the other in $\mathcal{H}_{B}$). The states having this property are called intricate, and we recall the connection with Sweedler's notation for the coproduct of a coalgebra \ref{sweedler}\label{bellpair}. 

 In addition to these spin properties, it is possible to experimentally create this pair in such a way that  the 2 spins $\frac{1}{2}$ go in 
opposite spatial directions.

\subsubsection{Measurements of a Bell pair}
\label{MBp}

We imagine the experiment where a Bell pair is emitted
at the origin of space and where the 2 elements of the pair propagate
in 2 opposite directions. Very far, Alice and Bob wait each one
for the arrival of their spin with their experimental apparatus
of spin measurement.

We imagine that Alice measures the first spin $\frac{1}{2}$
according to $A_{\theta_{1}}\equiv\sigma_{\theta_{1},0}\otimes I$ and that
Bob measures ``at the same time'' \footnote{The operators $A_{\theta_{1}}$ and $B_{\theta_{2}}$
been commuting, the precise order of the measurements is not important, as shown by the
show the formulas  (\ref{eq:M1},\ref{eq:M2}). }  the second spin $\frac{1}{2}$ according to $B_{\theta_{2}}\equiv I\otimes\sigma_{\theta_{2},0}$.
Applying the formula (\ref{eq:PEQ-1-1-1}), we find  

$
\mathbb{P}_{\left|Bell\right\rangle }\left(A_{\theta_{1}}=1,B_{\theta_2}=1\right) = \left|\left|Bell\right\rangle ^{+}\psi_{\theta_{1},0}\otimes\psi_{\theta_{2},0}\right|^{2}$
$$=\frac{1}{2}\left|\left(\left|-1\right\rangle \otimes\left|1\right\rangle -\left|1\right\rangle \otimes\left|-1\right\rangle \right)^{+}\left(\begin{array}{c}
\cos\left(\frac{\theta_{1}}{2}\right)
\sin\left(\frac{\theta_{1}}{2}\right)
\end{array}\right)\otimes\left(\begin{array}{c}
\cos\left(\frac{\theta_{2}}{2}\right)
\sin\left(\frac{\theta_{2}}{2}\right)
\end{array}\right)\right|^{2}$$
\begin{equation}
= \frac{1}{2}\left(\sin\frac{\theta_{1}}{2}\cos\frac{\theta_{2}}{2}-\cos\frac{\theta_{1}}{2}\sin\frac{\theta_{2}}{2}\right)^{2}=\frac{1}{2}\sin^{2}\left(\frac{\theta_{1}-\theta_{2}}{2}\right).\label{eq:Bell11}
\end{equation}

And likewise: 

\begin{equation}
\begin{cases}
\mathbb{P}_{\left|Bell\right\rangle }\left(A_{\theta_{1}}=-1,B_{\theta_{2}}=-1\right)=\left|\left|Bell\right\rangle ^{+}\omega_{\theta_{1},0}\otimes\omega_{\theta_{2},0}\right|^{2}=\frac{1}{2}\sin^{2}\left(\frac{\theta_{1}-\theta_{2}}{2}\right)\\
\mathbb{P}_{\left|Bell\right\rangle }\left(A_{\theta_{1}}=1,B_{\theta_{2}}=-1\right)=\left|\left|Bell\right\rangle ^{+}\psi_{\theta_{1},0}\otimes\omega_{\theta_{2},0}\right|^{2}=\frac{1}{2}\cos^{2}\left(\frac{\theta_{1}-\theta_{2}}{2}\right)\\
\mathbb{P}_{\left|Bell\right\rangle }\left(A_{\theta_{1}}=-1,B_{\theta_{2}}=1\right)=\left|\left|Bell\right\rangle ^{+}\omega_{\theta_{1},0}\otimes\psi_{\theta_{2},0}\right|^{2}=\frac{1}{2}\cos^{2}\left(\frac{\theta_{1}-\theta_{2}}{2}\right)
\end{cases}.\label{eq:Bell111}
\end{equation}

Noting $\mathbb{E}_{\left|Bell\right\rangle }$the expectation associated with the probability law $\mathbb{P}_{\left|Bell\right\rangle }$, this leads to

\begin{equation}
\mathbb{E}_{\left|Bell\right\rangle }\left(A_{\theta_{1}}B_{\theta_{2}}\right)=-\cos\left(\theta_{1}-\theta_{2}\right).\label{eq:magnifique}
\end{equation}

In particular, we find the perfect anti-correlation
\footnote{ This anti-correlation, together with the fact that Alice and Bob make their measurements
points in space as far away as they want, was for Einstein-Podolsky-Rosen an argument  in favor of a local common cause, whose existence would imply this anti-correlation but would also entail the
 incompleteness of the usual quantum mechanics. The missing variable
could be the fixation of the 2 spins just after the separation of the pair
 at the origin of space. Moreover, it is the lack of knowledge
of this hidden variable which would be the source of the randomness of quantum mechanics.
See the book of Laloë \cite{laloe} for many more ideas on this subject of hidden variables.} 
$\mathbb{E}_{\left|Bell\right\rangle }\left(A_{\theta_{1}}B_{\theta_{1}}\right)=-1$, 
expected from the form \eqref{etatpur} that is, if Alice and Bob query the system with the same measure, their results are anti-correlated, whatever the
measure ($\theta_{1}$) chosen. A Bell pair is then the metaphor
of a couple always in disagreement, whatever the (identical) question asked to both members of the couple! Bell's inequalities will focus on the statistics of the answers when they are asked different questions. 

We now imagine a more advanced experience than the previous one: 

\begin{itemize}

\item On the one hand, Alice still measures the first spin $\frac{1}{2}$, but
now with 2 possible choices of angle: $\theta_{1}$ or $\theta'_{1}$.
Alice notes the sequence of chosen angles and the sequence of associated results.
One can imagine that Alice did not choose in advance the order of the measurements
but that she uses for each measurement a balanced coin
to flip a coin: heads, she measures $\theta_{1}$; tails, she measures $\theta'_{1}$.

\item On the other side, Bob (potentially very very far from Alice but at equidistance with the place of separation of the pair) measures the other spin $\frac{1}{2}$
at the same time, with for him also 2 possible choices of spin measurement: $\theta_{2}$
or $\theta'_{2}$. Bob also notes his sequence of questions and the sequence
of associated results. Bob also chooses the measure that he realizes with
a coin toss with another balanced coin.

\item Then Alice and Bob get together (at least by phone) to compare their (very long) lists and make statistics.

\end{itemize}

We then define the Bell factor 

{\small
\begin{equation}
\mathcal{B}_{\theta_{1},\theta'_{1},\theta_{2},\theta'_{2}}\equiv\mathbb{P}_{\left|Bell\right\rangle }\left(A_{\theta'_{1}}=B{}_{\theta'_{2}}\right)+\mathbb{P}_{\left|Bell\right\rangle }\left(A_{\theta_{1}}=B{}_{\theta'_{2}}\right)+\mathbb{P}_{\left|Bell\right\rangle }\left(A_{\theta'_{1}}=B{}_{\theta_{2}}\right)-\mathbb{P}_{\left|Bell\right\rangle }\left(A_{\theta_{1}}=B{}_{\theta_{2}}\right).\label{eq:FB}
\end{equation}}

For example, $\mathbb{P}_{\left|Bell\right\rangle }\left(A_{\theta'_{1}}=B{}_{\theta'_{2}}\right)$
is the rate at which Alice measured her spin with $\sigma_{\theta'_{1},0},$
and Bob measured his spin with $\sigma_{\theta'_{1},0},$
and where Alice and Bob got the same result: both $+1$ or both $-1.$

The probabilities (\ref{eq:Bell11},\ref{eq:Bell111})
then give the quantum prediction $\mathbb{P}_{\left|Bell\right\rangle }\left(A_{\theta{}_{1}}=B{}_{\theta{}_{2}}\right)=\sin^{2}\left(\frac{\theta_{1}-\theta_{2}}{2}\right)$,
and we have the theoretical quantum expression of the Bell factor:
\[
\mathcal{B}_{\theta_{1},\theta'_{1},\theta_{2},\theta'_{2}}=\sin^{2}\left(\frac{\theta'_{1}-\theta'_{2}}{2}\right)+\sin^{2}\left(\frac{\theta_{1}-\theta'_{2}}{2}\right)+\sin^{2}\left(\frac{\theta'_{1}-\theta_{2}}{2}\right)-\sin^{2}\left(\frac{\theta_{1}-\theta_{2}}{2}\right).
\label{FB}
\]

In particular, we note that $\mathcal{B}_{\theta_{1},\theta'_{1},\theta_{2},\theta'_{2}}$
can be negative: we can have for example $\mathcal{B}_{0,\frac{2\pi}{3},\pi,\frac{\pi}{3}}=-\frac{1}{4}$. 

\subsection{The quest for a classical probability space representing quantum probabilities}

 \subsubsection{Bell factor in  classical probability}
With the situation described in the previous paragraph, we can ask ourselves the question of the existence of a model of this experiment by classical probabilities. To do this, let us start by stating the natural hypothesis:
 
\textbf{\textit{Hypothesis $\mathcal{C}$ }} :  It is assumed that there exists a classical probability space
where the 4 classical random variables
$A_{\theta{}_{1}},A_{\theta'_{1}},B_{\theta{}_{2}},B_{\theta'_{2}}$,\textcolor{black}{{} taking values in $\pm1$,} have a joint law consistent with the marginals given by the
quantum expressions(\ref{eq:Bell11},\ref{eq:Bell111}). 

\vspace{0.5cm}

In the logic of variables with $\pm1$ values, we can see that we necessarily have the inclusion

\begin{equation}
\left\{ A_{\theta{}_{1}}=B_{\theta{}_{2}}\right\} \in\left(\left\{ A_{\theta{}_{1}}=B_{\theta'_{2}}\right\} \cup\left\{ A_{\theta'_{1}}=B_{\theta'_{2}}\right\} \cup\left\{ A_{\theta'_{1}}=B_{\theta{}_{2}}\right\} \right).\label{eq:log}
\end{equation}

For example: If I am in the set $A_{\theta{}_{1}}=B_{\theta{}_{2}}=1,$
and if I do not want to be in the set $A_{\theta{}_{1}}=B_{\theta'_{2}},$
then necessarily $B_{\theta'_{2}}=-1$. But if in addition I do not want
not to be in $\left\{A_{\theta'_{1}}=B_{\theta'_{2}}\right\}$,
then necessarily $A_{\theta'_{1}}=1$. And thus I am necessarily in
the set $\left\{ A_{\theta'_{1}}=B_{\theta_{2}}\right\} .$

\vspace{0.5cm}

Then, under the \textit{Hypothesis $\mathcal{C}$}, we have therefore necessarily positivity
of the Bell factor (\ref{eq:FB}) $\mathcal{B}_{\theta_{1},\theta'_{1},\theta_{2},\theta'_{2}}$.
Now we have shown in the previous paragraph that quantum probabilities
allow $\mathcal{B}_{0,\frac{2\pi}{3},\pi,\frac{\pi}{3}}=-\frac{1}{4}.$
The Orsay experiments by Alain Aspect and colleagues \cite{aspect}
have confirmed experimentally the possibility of negative values for $\mathcal{B}_{\theta_{1},\theta'_{1},\theta_{2},\theta'_{2}}$.  

\vspace{0.5cm}

One is thus obliged to conclude that the \textbf{\textit{Hypothesis
}}\textit{$\mathcal{C}$}\textbf{ is false in quantum mechanics}:
there is no classical probability space where the 4 classical
 random variables $A_{0},A_{\frac{2\pi}{3}},B_{\pi},B_{\frac{\pi}{3}}$,
taking values in $\pm1$, have a joint law consistent with the marginals (\ref{eq:Bell11},\ref{eq:Bell111}) given by
the quantum rules. However, there are classical joint laws
 (\ref{eq:Bell11},\ref{eq:Bell111}) for the pairs $\left(A_{0},B_{\pi}\right),\left(A_{0},B_{\frac{\pi}{3}}\right),\left(A_{\frac{2\pi}{3}},B_{\pi}\right)$
and $\left(A_{\frac{2\pi}{3}},B_{\frac{\pi}{3}}\right)$.

Note
that in practice, at each joint arrival of the two spins, it is realized either the measurement $\left(A_{0},B_{\pi}\right)$,
or (exclusive) the measure $\left(A_{0},B_{\frac{\pi}{3}}\right)$, or (exclusive) $\left(A_{\frac{2\pi}{3}},B_{{\pi}}\right)$, and finaly or (exclusive) $\left(A_{\frac{2\pi}{3}},B_{\frac{\pi}{3}}\right)$.  You can't do better at a given time and measure at the same time
time $A_{0},A_{\frac{2\pi}{3}},B_{\pi},B_{\frac{\pi}{3}}.$ 

We are thus obliged a priori to give up considering  the results of experiments,  that we could potentially do but that we have not yet done,  as classical random variables. 
This type of results is at the origin of multiple debates on the general meaning of quantum mechanics and, more specifically, on the interpretation to be given to  such results. 

In fact, one could consider the same
escape as in the paragraph (\ref{subsec:arg1}). That is to say, consider that the random variable translating the measurement
of the spin by Bob depends (i.e. is influenced) by the angle chosen
by Alice (i.e. $B_{\theta_{2}}(\theta_{1})$ ), and vice versa.
This would still invalidate the inclusion (\ref{eq:log}). But this time it would imply a break
of locality due to the potentially infinite distance between Alice and Bob, and locality \footnote{ 
This statement is not universally accepted.
For example, Bohmians consider that locality is necessarily broken\cite{bricmont}. }.  

\subsubsection{The unreasonable efficiency
of the quantum probabilities}

For pedagogical purposes, it is interesting to reformulate the previous experiment in the context of a
 card game \cite{maasen}  with a classical version where we use two balanced coins
and a quantum version where we use a Bell pair.

\begin{itemize}

\item Classical version of the game with two balanced coins: 

\begin{itemize}
\item 2 players Alice and Bob agree on a game strategy, 
then separate and can no longer communicate at all. This
strategy can be random and depends on the outcome of 2 
balanced coins they each have in their possession.

\item 1 referee, who iteratively and randomly distributes black or red cards
 to Alice and Bob. Each player sees only his own
card.

\item Alice and Bob can throw their respective coins before answering.
Each sees only his own coin. 

\item Based on their cards, on the results of their coin tosses and on their predefined strategy, Alice and Bob must simultaneously say Yes or No. We
call $A_{R}$, $A_{N}$, $B_{R}$ and $B_{N}$ their respective answers according to the color of their card. 

\item The cards are then turned over and the referee writes down $0$ (if their answer is different) or $1$ (if their answer is the same), in a table of 4 boxes according to the colors of the cards: $RR,RN,NR,NN.$

\item Goal of the game: choose a strategy such that, after a very large number
of iterations, the proportion of $1$ in the square $RR$ is greater than the sum of the 3 other squares.
More precisely, the goal is to try to have
$$\mathbb{P}\left(A_{R}=B_{R}\right)>\mathbb{P}\left(A_{N}=B_{N}\right)+\mathbb{P}\left(A_{N}=B_{R}\right)+\mathbb{P}\left(A_{R}=B_{N}\right).$$
\end{itemize}
\vspace{0.5cm}
There are two implicit, but crucial, assumptions in the rule of this game, which translate two central ideas in physics: that there is no effect without a cause (what philosophers sometimes call ``the principle of reason"); the other that there is no instantaneous action at a distance,

\vspace{0.5cm}

\begin{center}
\textbf{($\mathcal{HR}$) Hypothesis} of \textbf{\textit{reality}}: the strategy (even random) of the players is fixed at the beginning and pre-exists
to the realization of the experiment.

\par
\end{center}

\vspace{0.5cm}

\begin{center}
\textbf{($\mathcal{HL}$) Hypothesis} of \textbf{\textit{locality}}:
once separated, Alice and Bob cannot communicate in any way.
Neither the card Alice receives, nor the face of her coin, influences Bob's answer, and vice versa. 
\par
\end{center}
\vspace{0.5cm}

We have seen previously that for all classical probability laws
of bivalent random variables $\pm1$, we have the Bell inequality

\[
\mathbb{P}\left(A_{R}=B_{R}\right)\leq \mathbb{P}\left(A_{N}=B_{N}\right)+\mathbb{P}\left(A_{R}=B_{N}\right)+\mathbb{P}\left(A_{N}=B_{R}\right).
\]
So, Alice and Bob can't reach the goal
of Bell's game! We can still obtain a predefined strategy where
we get the case of equality in the previous inequality. This is the
strategy where, no matter what the value of their coin toss is, Alice always says
``Yes'' so 
$\begin{cases}
A_{R}=Yes\\
A_{N}=Yes
\end{cases}$ 
and Bob answers ``Yes" if and only if his card is red,  so
$\begin{cases}
B_{R}=Yes\\
B_{N}=No
\end{cases}$.
We have then 
 $
 \begin{cases}
\mathbb{P}\left(A_{R}=B_{R}\right)=1\\
\mathbb{P}\left(A_{R}=B_{N}\right)=0\\
\mathbb{P}\left(A_{N}=B_{R}\right)=1\\
\mathbb{P}\left(A_{N}=B_{N}\right)=0
\end{cases}, 
$ and  $$1=\mathbb{P}\left(A_{R}=B_{R}\right)=\mathbb{P}\left(A_{N}=B_{N}\right)+\mathbb{P}\left(A_{R}=B_{N}\right)+\mathbb{P}\left(A_{N}=B_{R}\right).$$

\item ``Quantum" version of the game with a Bell's pair: 

The game is identical to the ``classic" version except that we replace the 2 coins by a pure state of Bell's pair, and that :

\begin{itemize}

\item If Alice receives a red (resp. black) card from the referee, then she waits for her spin and
 measures $A_{\theta_{1}}\equiv\sigma_{\theta_{1},0}\otimes I$
(resp $A_{\theta'_{1}}\equiv\sigma_{\theta'_{1},0}\otimes I$ ). 

\item If Bob receives a red (resp. black) card from the referee, then he waits for his spin and
 measures $B_{\theta_{2}}\equiv I\otimes\sigma_{\theta_{2},0}$
(resp $B_{\theta'_{2}}\equiv I\otimes\sigma_{\theta'_{2},0}$ ). 

\end{itemize}

Finally, the predefined strategy set by Alice and Bob is that they answer ``Yes'' when they get the $+1$ measurement result
and ``No'' when they get the measurement result $-1$. 

We have then 
\[
\mathbb{P}\left(A_{\theta_{1}}=B_{\theta_{2}}\right)-\mathbb{P}\left(A_{\theta'_{1}}=B_{\theta'_{2}}\right)-\mathbb{P}\left(A_{\theta'_{1}}=B_{\theta_{2}}\right)-\mathbb{P}\left(A_{\theta_{1}}=B_{\theta'_{2}}\right)=-\mathcal{B}_{\theta_{1},\theta'_{1},\theta_{2},\theta'_{2}}.
\]

Incredible..... We can reach the goal of Bell's game by using
the same values of measures as in the paragraph \ref{MBp}, 
because $\mathcal{B}_{0,\frac{2\pi}{3},\pi,\frac{\pi}{3}}=-\frac{1}{4}$
and then 
\[
\mathbb{P}\left(A_{R}=B_{R}\right)=\frac{1}{4}+\left(\mathbb{P}\left(A_{N}=B_{N}\right)+\mathbb{P}\left(A_{N}=B_{R}\right)+\mathbb{P}\left(A_{R}=B_{N}\right)\right).
\]
The quantum reaches the goal of the game! We are forced to conclude that \textbf{in quantum necessarily at least one of the two hypotheses is broken} \footnote{\label{footnote} 
Which hypothesis is broken is not universally accepted.
For example, the Bohmians consider that it is necessarily the hypothesis
of \textit{locality}($\mathcal{HL}$)
which is broken. See \cite{bricmont} for more on this approach 
and \cite{werner}
for a very clear critique of this view.
Conversely, some physicists,
for example C. Tresser \cite{tresser}
and M. Zukowski and \v{C} Brukner \cite{zuko}
consider that it is necessarily the hypothesis of
\textit{reality} ($\mathcal{HR}$)
which is broken.  }
\textbf{between the hypothesis of} \textbf{\textit{\textcolor{black}{reality}}}\textbf{\textcolor{black}{{}
}}\textbf{($\mathcal{HR}$)}\textbf{\textcolor{black}{{} and the hypothesis of }}\textbf{\textit{\textcolor{black}{locality }}}\textbf{($\mathcal{HL}$)}\textbf{\textcolor{black}{{}
! }}
\end{itemize}

\vspace{1cm}

\subsubsection{The abyss of interpretation}

Since its birth in the 1920's, the quantum theory has always been associated with a certain strangeness, because it seems to go against the intuition
of the real world. This strangeness is only reinforced by the brief study of Bell's inequalities
 that we have presented. In a rather philosophical way,
we can deduce from Bell's inequalities that the measured values do not
do not pre-exist the measurement, and that the measurement does not have the effect of
revealing a pre-existing reality. Some consider that this amounts to
to renounce the notion of \textit{realism}, others to the notion of
\textit{factuality}. Let us note that this interpretation was strongly
opposed by \textcolor{black}{Albert Einstein who said: }

\begin{center}
\textcolor{black}{\textquotedblleft }\textit{\textcolor{black}{I like
to think that the moon is there even if I am not looking at it }}\textcolor{black}{\textquotedblright . }
\par
\end{center}

We refer to the book of Laloë \cite{laloe}
for an enlightened panorama in an enormous literature, often at the borderline with
philosophy, which tries to interpret all this. We will close
this part with an amusing quote from Frohlich and Schnubel  \cite{Frohlich}
\begin{center}
\textit{``The ``foundations of quantum mechanics" represent a notoriously thorny and enigmatic subject. Asking twenty-five grown up physicists to present their views on the foundations
of quantum mechanics, one can expect to get the following spectrum of reactions : Three
will refuse to talk – alluding to the slogan ``shut up and calculate" – three will say that the
problems encountered in this subject are so difficult that it might take another 100 years
before they will be solved; five will claim that the ``Copenhagen Interpretation", has
settled all problems, but they are unable to say, in clear terms, what they mean; three will
refer us to Bell’s book (but admit they have not understood it completely); three confess to
be ``Bohmians" (but do not claim to have had an encounter with Bohmian trajectories);
two claim that all problems disappear in the Dirac-Feynman path-integral formalism;
another two believe in ``many worlds" but make their income in our’s, and two advocate
``consistent histories"; two swear on QBism , (but have never seen ``les demoiselles
d’Avignon"); two are convinced that the collapse of the wave function - spontaneous or
not - is fundamental; and one thinks that one must appeal to quantum gravity to arrive at a
coherent picture.
Almost all of them are convinced that theirs is the only sane point of view. Many workers
in the field have lost the ability to do technically demanding work or never had it...''}.
\end{center}



\newpage


\begin{thebibliography}{mmmm}

\bibitem{AG} Agrachev, A. A.,  Gamkrelidze, R. V. Chronological algebras and nonstationary vector fields. Itogi Nauki Tekh., Ser. Probl. Geom. 11 (1980), 135-176. In \textit{Russian. English transl., J. Sov. Math.} 17, 1650–1675 (1981).
\bibitem{Akemann}, G.Akemann, J.Baik,  P. Di Francesco.  The Oxford Handbook of Random Matrix Theory. Oxford: Oxford University Press (2011). 
\bibitem{aldous}  D. Aldous, P. Diaconis, Shuffling Cards and Stopping Times, The American Mathematical Monthly, 93:5,  (1986) 333-348.
\bibitem{Anderson} G.W. Anderson, A. Guionnet, O. Zeitouni, An introduction to random matrices. Cambridge: Cambridge University Press. (2010)
\bibitem{A1} M. Anshelevich,  Appell polynomials and their relatives. Internat.Math. Res.Notices,  65, (2004) 3469-3531.
\bibitem{A2} M. Anshelevich, Appell polynomials and their relatives. ii: Boolean theory. Indiana Univ. Math. J.,
58(2), (2009), 929-968.
\bibitem{A3} M. Anshelevich, Appell polynomials and their relatives. iii: Conditionally free theory. Illinois J.
Math., 53(1), (2009) 39-66.
\bibitem{Bohigas}
O. Bohigas, M.J. Giannoni, Schmit, Characterization of Chaotic Quantum Spectra and Universality of Level Fluctuation Laws. Phys. Rev. Lett. 52 (1),(1984), 1–4. 
\bibitem{lehner-etal_15}
O.~Arizmendi, T.~Hasebe, F.~Lehner, C.~Vargas,
{\it{Relations between cumulants in noncommutative probability}},
Adv.~Mathematics {\bf{282}}, (2015) 56-92.
\bibitem{aspect} A.
Aspect, P. Grangier, G. Roger, Experimental realization of Einstein-Podolsky-Rosen-Bohm
Gedankenexperiment: A new violation of Bell's inequalities, in
Physical Review Letters, Vol. 49, no 2, p. 91 - 94 1982.
\bibitem{bauer} M. Bauer, R. Chetrite, K. Ebrahimi-Fard, F. Patras, Time-ordering and a generalized Magnus expansion, Letters Math. Phys. (2013), 103 (3), 331-350. 
\bibitem{bell}J.S. Bell, On
the Einstein Podolsky Rosen Paradox, Physics,
1 (3): 195 - 200, 1964
\bibitem{bricmont}J. Bricmont, Making sense of quantum mechanics (Vol. 37). Berlin: Springer, 2016.
\bibitem{bricmont2} J. Bricmont: La mécanique quantique pour non-physiciens, cours UCL-FYMA, Louvain, 2018.
\bibitem{CP21}
P.~Cartier, F.~Patras,
{\it{Classical Hopf algebras and their Applications}},
Berlin Heidelberg, Springer, 2021.
\bibitem{cepp}A. Celestino, K. Ebrahimi-Fard, F. Patras,
D. Perales, Cumulant--Cumulant Relations in Free Probability Theory
from Magnus’ Expansion, Foundations of Computational Mathematics (2021), 1-23.
\bibitem{CelP22} A. Celestino, F. Patras, A forest formula for pre-Lie exponentials, Magnus' operator and cumulant-cumulant relations. arXiv:2203.1196810. 
\bibitem{CP13} F. Chapoton, F. Patras, Enveloping algebras of preLie algebras, Solomon idempotents and the Magnus formula, International Journal of Algebra and Computation 23.04 (2013) 853-861.
\bibitem{dpr} P. Diaconis, C.Y. Pang, A. Ram, Hopf algebras and Markov chains: two examples and a theory, J. Alg. Comb.  39 (3) (2014) 527-585.
\bibitem{persi} P. Diaconis,  From shuffling cards to walking around the building: An introduction to modern Markov chain theory. Doc. Math. (Bielefeld) Extra Vol. ICM Berlin, 1998, vol. I. pp. 187–204.
\bibitem{ebrahimipatras_15}
K.~Ebrahimi-Fard, F.~Patras,
{\emph{Cumulants, free cumulants and half-shuffles}}, 
Proc.~R.~Soc.~A {\bf{471}}, 2176, (2015).
\bibitem{EP16} K. Ebrahimi-Fard, F. Patras, The splitting process in free probability theory. \textit{Int. Math. Res. Not.} 9,
2647-2676 (2016).
\bibitem{EFP16} K. Ebrahimi-Fard, F. Patras, The combinatorics of Green's functions in planar field theories. Front. Phys. 11(6), 110310 (2016).
\bibitem{EFP18} K. Ebrahimi-Fard, F. Patras, Monotone, free, and boolean cumulants from a Hopf algebraic point of view. \textit{Adv. Math.} 328, 112-132 (2018).
\bibitem{EFP19} K. Ebrahimi-Fard, F. Patras, Shuffle group laws. Applications in free probability. \textit{P. Lond. Math.
Soc.} 119, 814-840 (2019).
\bibitem{efp22} K. Ebrahimi-Fard, F. Patras, From iterated integrals and chronological calculus to Hopf and Rota-Baxter algebras. Algebra and Applications 3, Combinatorial algebras and Hopf algebras, coordinated by A. Makhlouf, ISTE Ltd-Wiley (to appear).
\bibitem{eptz0} K. Ebrahimi-Fard, F. Patras, N. Tapia,
L. Zambotti, Hopf-algebraic Deformations of Products and Wick Polynomials. International Mathematics Research Notices, rny269, pp. 1–36, 2018. 
\bibitem{eptz} K. Ebrahimi-Fard, F. Patras, N. Tapia,
L. Zambotti, Wick polynomials in noncommutative
probability: a group-theoretical approach, Canad. J. Math. 2021, pp. 1-27.
\bibitem{epr}A. Einstein,
B. Podolsky, N. Rosen, Can Quantum-Mechanical Description
of Physical Reality Be Considered Complete? , Phys. Rev., vol. 47,
1935, p. 777-780
 \bibitem{feynman} R. Feynman,
 R. B. Leighton, The Feynman
Lectures on Physics, Vol. 3. Addison-Wesley, Matthew Sands, 1965.
\bibitem{friedrich} R. Friedrich and J. McKay, Homogeneous Lie groups and quantum probability,
arXiv:1506.07089v1
 \bibitem{Frohlich}  J. Frohlich, B. Schubnel, 
Quantum probability theory and the foundations of quantum mechanics. 
The Message of Quantum Science, 131-193, 2015. 
\bibitem{Gelfand} Gelfand, I. M., Krob, D., Lascoux, A., Leclerc, B., Retakh, V. S. and Thibon, J.-Y. (1995). Non-
commutative symmetric functions. Adv. in Math., 112, 218–348
\bibitem{Ger}M. Gerstenhaber, The cohomology structure of an associative ring, Annals of Mathematics (1963) 267--288.
\bibitem{gleason}A. M. Gleason, Measures on the closed subspaces
of a Hilbert space . Indiana University Mathematics
Journal. 6 (4) (1957) 885-893.
\bibitem{Lehner}T. Hasebe, F. Lehner, Cumulants, Spreadability and the Campbell-Baker-Hausdorff Series, https://arxiv.org/abs/1711.00219.
\bibitem{hasebesaigo_11}
T.~Hasebe, H.~Saigo
{\emph{The monotone cumulants}}, 
Annales de l'Institut Henri Poincar\'e - Probabilit\'es et Statistiques {\bf{47}}, No.~4, (2011) 1160-1170.
\bibitem{heisenberg} W. Heisenberg, Über den anschaulichen Inhalt der quantentheoretischen Kinematik und Mechanik. Z. Physik 43, 172–198 (1927).
\bibitem{holevo} A. S. Holevo, Statistical Structure of Quantum Theory, Springer, 2001.
\bibitem{Hrz22} Hruza, L., Bernard, D. (2022). Dynamics of Fluctuations in the Open Quantum SSEP and Free Probability. arXiv preprint arXiv:2204.11680.
\bibitem{joos} E. Joos, H. D. Zeh, C.
Kiefer, D. Giulini, J. Kupsch, I.-O. Stamatescu, Decoherence and the Appearance of
a Classical World in Quantum Theory , 
Springer 2003.
\bibitem{kolmogorov} A. Kolmogorov,
 Foundations of the theory of probability, New York, USA:
Chelsea Publishing Company, 1933.
\bibitem{LM} T. Lada, M. Markl, Symmetric brace algebras, Applied Categorical Structures 13.4 (2005) 351-370.
 \bibitem{laloe} F. Laloë,
Comprenons-nous vraiment la m\'ecanique Quantique ? , CNRS éditions, 2ième éd., 2018. 
\bibitem{lan} L.D. Landau, The Damping
Problem in Wave Mechanics (1927) in Collected Papers
of L.D. Landau. 1965. pp. 8-18.
\bibitem{Laz} M. Lazard, Lois de groupes et analyseurs,
Annales scientifiques de l'\'Ecole Normale Sup\'erieure, S\'erie 3, Tome 72 (1955) no. 4, p. 299-400.
\bibitem{GO} J.-M. Oudom, D. Guin, On the Lie enveloping algebra of a pre-Lie algebra, Journal of K-theory 2.1 (2008) 147-167.
\bibitem{Lenc}R. Lenczewski,  Quantum central limit theorems. In Symmetries in Science VIII (pp. 299-314). Springer, Boston, 1995.
\bibitem{maasen} H. Maasen,  Quantum Probability and Quantum Information Theory, Quantum information, computation and cryptography, 65-108, 2010.
\bibitem{manzel} S. Manzel, M. Schürmann, Non-commutative stochastic independence and cumulants, Infin. Dimens. Anal. Quantum Probab. Relat. Top. 20 (2017) 1750010.
\bibitem{Mehta} M.L. Mehta  Random Matrices. Amsterdam: Elsevier/Academic Press. (2004)
\bibitem{Men} F. Menous, F. Patras, Right-handed Hopf algebras and the preLie forest formula. Annales de l'Institut Henri Poincar\'e D, 5(1), (2018) 103-125.
\bibitem{mermin}D.
Mermin, Is the moon there when nobody looks. Reality and the quantum
theory, Phys. Today, April 1985 
\bibitem{Mingo} J. Mingo, R. Speicher.  Free Probability and Random Matrices 
Fields Institute Monographs, Springer, (2017)
\bibitem{MPl} Mielnik, B., Pleba\'nski, J. Combinatorial approach to Baker-Campbell-Hausdorff exponents. In Annales de l'IHP Physique th\'eorique Vol. 12, No. 3, (1970) pp. 215-254.
\bibitem{Mura03} N. Muraki, The five independences as natural products, Infinite Dimensional Analysis, Quantum Probability and Related TopicsVol. 06, No. 03, pp. 337-371 (2003).
\bibitem{Murua}A. Murua, The Hopf algebra of rooted trees, free Lie algebras, and Lie series, Found. Comput.Math. 6,
387-426 (2006).
\bibitem{NS} Neu P., Speicher R., A self-consistent master equation and a new kind of cumulants. Zeitschrift für Physik B Condensed Matter, 92(3), (1993) 399-407.
\bibitem{nicaspeicher_06}
A.~Nica, R.~Speicher, 
Lectures on the combinatorics of free probability,
London Mathematical Society Lecture Note Series, {\bf{335}} Cambridge University Press (2006).
\bibitem{Pap22} Pappalardi, S., Foini, L., Kurchan, J. (2022). Eigenstate Thermalization Hypothesis and Free Probability. arXiv preprint arXiv:2204.11679.
\bibitem{pat92} F. Patras, Homoth\'eties simpliciales. PhD thesis, University Paris 7, January 1992.
\bibitem{pat93} F. Patras, La décomposition en poids des algèbres de Hopf. Ann. Inst. Fourier. 43, 4 (1993), 1067-1087.
\bibitem{pat94} F. Patras, L'algèbre des descentes d'une bigèbre graduée. J. Algebra 170, 2 (1994), 547-566.
\bibitem{pp} Patras, F. et Planas-Bielsa, V., Complex Systems: From the Presocratics to Pension Funds, in Complexity and Emergence, Springer 2022 (à paraître).
\bibitem{peccati} G. Peccati, M. S. Taqqu, Wiener Chaos: Moments,
Cumulants and Diagrams.
A survey with computer implementation, Springer, 2011.
\bibitem{peres2} A. Peres, Unperformed experiments have no results,  American Journal of Physics 46, 745 (1978)
\bibitem{peres} A. Peres, Quantum
Theory, Concepts and Methods. Kluwer academic, 1993.
\bibitem{Poincare}Poincar{\'e} H., Calcul des probabilit{\'e}s, Gauthier-Villars, 1912
\bibitem{Reutenauer} 
C.~Reutenauer, 
Free Lie algebras, 
Oxford University Press (1993).
\bibitem{Schu} M. Sch\"urmann, A central limit theorem for coalgebras, in Probability measures
on groups VIII, Proceedings, Oberwolfach 1985, Ed. Heyer H., Leet. Notes in Math.
1210, 153-157, Springer 1986.
\bibitem{Spei1} R. Speicher, A new example of ``Independence" and ``White Noise", Probab.
Th. ReI. Fields 84, 141-159 (1990).
\bibitem{Spei2} R. Speicher, A non-commutative central limit theorem, Math. Z. 209, 55-66 (1992).
\bibitem{Spe94} Speicher, R. Multiplicative functions on the lattice of noncrossing partitions and free convolution.
\textit{Math. Ann.} 298 no. 4, 611–628 (1994).
\bibitem{speicher_97c} 
R.~Speicher, R.~Woroudi,
{\it{Boolean convolution}},
In: Voiculescu, D. V. (ed.) Free Probability Theory. Proceedings, Toronto, Canada 1995, 
Fields Inst.~Commun.~{\bf{12}}, Providence, RI: Amer.~Math.~Soc., (1997) 267-279.
\bibitem{tresser} C. Tresser, Bell's theory with no locality assumption. Eur. Phys. J.D 2010.
\bibitem{Vin} E. Vinberg, The theory of homogeneous convex cones, Trudy Moskovskogo Matematicheskogo Obshchestva 12 (1963) 303-358; English transl. The theory of convex homogeneous cones, Moscow Math. Soc. 12 (1963) 340-403.
\bibitem{voiculescu_92} 
D.~Voiculescu, K.~J.~Dykema, A.~Nica, 
Free random variables, 
CRM Monograph Series, vol.~1, American Mathematical Society, Providence, RI, 1992. 
A noncommutative probability approach to free products with applications to random matrices, operator algebras and harmonic analysis on free groups.
\bibitem{voiculescu_95} 
D.~Voiculescu, 
{\it{Free Probability Theory: Random Matrices and von Neumann Algebras}}, 
Proceedings of the International Congress of Mathematicians, Z\"urich, 
Switzerland 1994. Birkh\"auser Verlag, Basel, Switzerland (1995).
\bibitem{vN} J. von Neumann, Wahrscheinlichkeitstheoretischer Aufbau der Quantenmechanik,
Göttinger Nachrichten, 1 (1927) 245 - 272.
\bibitem{vWal} W. von Waldenfels, An algebraic central limit theorem in the anticommuting
case, Z. Wahr. Verw. Gebiete 42, , 135-140 (1979).
\bibitem{werner}R. Werner, Comment on `What Bell did', J Phys A 47 (2014) 424011.
\bibitem{Wick}G. C. Wick, The Evaluation of the Collision Matrix, Phys. Rev. 80, (1950) 268.
\bibitem{wigner} E. Wigner, The
Unreasonable Effectiveness of Mathematics in the Natural Sciences, Communications on Pure and Applied Mathematics
13, 1960.
\bibitem{Wigner2} E. Wigner, Characteristic vectors of bordered matrices with infinite dimensions.  Annals of Mathematics. 62 (3): 548–564. (1965)
\bibitem{wiseman}H. Wiseman, G. Milburn, Quantum Measurement and Control. Cambridge: Cambridge University
Press, 2009.
 \bibitem{Zeidler} E. Zeidler, Quantum Field Theory I: Basics in Mathematics and Physics. A Bridge between Mathematicians and Physicists, Springer, 2008.
\bibitem{zuko} M. Zukowski et \v{C}. Brukner, Quantum non-locality\textemdash it ain't necessarily so... Journal of Physics A: Mathematical and Theoretical 47 (42), 2014.
\end{thebibliography}
\end{document}